\documentclass[11pt, letterpaper]{article}

\usepackage[margin=1.2in, top=1.1in, bottom=1.1in]{geometry} 
\usepackage[T1]{fontenc}
\usepackage[utf8]{inputenc}
\usepackage{tgpagella} 
\usepackage[varg]{newtxmath} 
\usepackage{microtype} 

\usepackage{amsmath}
\usepackage{amsthm}
\usepackage{mathtools}
\usepackage{bm}

\usepackage{graphicx}
\usepackage{subcaption}
\graphicspath{ {Figures/} }
\usepackage{booktabs}  
\usepackage{array}
\newcolumntype{P}[1]{>{\centering\arraybackslash}p{#1}}
\usepackage{enumerate} 

\usepackage{xcolor}
\definecolor{DesignBlue}{RGB}{26, 54, 93}   
\definecolor{DesignGray}{RGB}{100, 110, 120} 

\usepackage{titlesec}
\titleformat{\section}{\large\bfseries\color{DesignBlue}}{\thesection.}{0.5em}{}
\titleformat{\subsection}{\normalsize\bfseries\color{DesignBlue}}{\thesubsection}{0.5em}{}
\titleformat{\subsubsection}{\normalsize\itshape\color{DesignBlue}}{\thesubsubsection}{0.5em}{}

\titlespacing*{\section}{0pt}{2.5ex plus 1ex minus .2ex}{1.5ex plus .2ex}

\usepackage{fancyhdr}
\usepackage{authblk}

\usepackage{cite}
\usepackage[psdextra]{hyperref}
\hypersetup{
    colorlinks=true,
    linkcolor=DesignBlue,
    citecolor=DesignBlue,
    urlcolor=DesignBlue
}
\usepackage{cleveref}

\newtheoremstyle{modern}
  {12pt}
  {12pt}
  {}
  {}
  {\bfseries\color{DesignBlue}}
  {.}
  {.5em}
  {}
  
\newtheoremstyle{modernremark}
  {6pt}{6pt}{\normalfont}{}{\itshape\color{DesignBlue}}{.}{.5em}{}

\theoremstyle{modern}
\newtheorem{thm}{Theorem}

\newtheorem{lem}{Lemma}

\newtheorem{define}{Definition}
\newtheorem{assume}{Assumption}

\theoremstyle{modernremark}
\newtheorem{rem}{Remark}

\Crefname{thm}{Theorem}{Theorems}
\Crefname{prop}{Proposition}{Propositions}
\Crefname{assume}{Assumption}{Assumptions}
\Crefname{problem}{Problem}{Problems}
\Crefname{lem}{Lemma}{Lemmas}

\usepackage{tikz}
\usetikzlibrary{shapes, arrows, positioning}
\tikzstyle{block} = [draw, rectangle, fill=white, minimum height=2.5em, minimum width=4em, line width=0.8pt]
\tikzstyle{sum} = [draw, circle, inner sep=0pt, minimum size=6mm, line width=0.8pt]

\def \ba{\begin{array}}
\def \ea{\end{array}}
\def \be{\begin{equation}}
\def \ee{\end{equation}}

\def \colsep{\arraycolsep}

\def \bb{\mathbb}

\def \mc{\mathcal}

\def \ol{\overline}

\newcommand\eqdef\coloneqq

\DeclareMathOperator*{\eig}{eig}
\DeclareMathOperator*{\rank}{rank}

\DeclareMathOperator*{\cond}{cond}

\newcommand{\keywords}[1]{%
    \vspace{1em}
    \noindent\textbf{\color{DesignBlue}Keywords---}%
    \hspace{0.5em}\parbox[t]{\dimexpr\textwidth-7em}{\raggedright #1}
}

\title{\vspace{-2em}\huge\bfseries\color{DesignBlue}
KKL Observer Synthesis for Nonlinear Systems via Physics-Informed Learning
}

\author[1]{M.~Umar~B.~Niazi}
\author[2]{John~Cao}
\author[1,3]{Matthieu~Barreau}
\author[1,3]{Karl~H.~Johansson}

\affil[1]{Department of Decision and Control Systems, EECS, KTH Royal Institute of Technology, SE-100 44 Stockholm, Sweden.} 
\affil[2]{Department of Engineering Science, University of Oxford, Parks Road, Oxford, OX1 3PJ, UK.} 
\affil[3]{Digital Futures, KTH Royal Institute of Technology, SE-100 44 Stockholm, Sweden.}

\date{}

\begin{document}
\maketitle
\thispagestyle{empty}

\begin{abstract}
    This paper proposes a novel learning approach for designing Kazantzis-Kravaris or nonlinear Luenberger (KKL) observers for autonomous nonlinear systems. The design of a KKL observer involves finding an injective map that transforms the system state into a higher-dimensional observer state, whose dynamics is linear and stable. The observer's state is then mapped back to the original system coordinates via the inverse map to obtain the state estimate. However, finding this transformation and its inverse is quite challenging. We propose learning the forward mapping using a physics-informed neural network, and then learning its inverse mapping with a conventional feedforward neural network. Theoretical guarantees for the robustness of state estimation against approximation error and system uncertainties are provided, including non-asymptotic learning guarantees that link approximation quality to finite sample sizes. The effectiveness of the proposed approach is demonstrated through numerical simulations on benchmark examples, showing better generalization capability outside the training domain compared to state-of-the-art methods.
\end{abstract}


\section{Introduction}

Observers are crucial in output feedback control \cite{teel1995}, fault diagnosis \cite{chen2007}, and digital twins \cite{dettu2024}. Accurate state estimation enables effective control, monitoring, and decision-making in various engineering domains. Full-state measurement of real-world systems is often impractical, or even impossible, due to limitations in sensing resources and capabilities. This necessitates the use of observers that employ available sensor measurements and the system's model to estimate the state. 

Several observer design methods, such as Luenberger-like observers, extended Kalman filters, and high-gain observers, have been proposed in the literature, each offering its own strengths and weaknesses depending on the specific application and system characteristics \cite{gauthier2001, khalil2014, bernard2019, boutat2021}. In this paper, we focus on Kazantzis-Kravaris/Luenberger (KKL) observers due to their applicability to a very general class of nonlinear systems and the well-developed theoretical framework for their design and analysis \cite{kazantzis1998, andrieu2006, andrieu2014, bernard2022, brivadis2023}.

Despite the advancements, synthesizing KKL observers remains challenging. Existing techniques assume knowledge of an injective map that transforms a nonlinear system into a special form required by the KKL observer. However, analytically finding this map is inherently difficult. Moreover, even when this map is known, as for some systems \cite{bernard2018}, computing its inverse to recover the state in the original, physically meaningful coordinates is also challenging \cite{andrieu2021}. This highlights the need for novel approaches that can overcome these limitations.

In this paper, we present a novel learning approach for designing KKL observers for autonomous nonlinear systems. At the core of KKL observer design is an injective coordinate transformation that lifts the nonlinear system to a higher-dimensional space. The transformed system exhibits two key properties: it is bounded-input-bounded-state stable and linear up to output injection.
The KKL observer operates by replicating the transformed system in the higher-dimensional space. To reconstruct the state estimate in the original coordinates, the left inverse of the transformation map is applied to the observer's state. The transformation map's injectivity property guarantees the accuracy of the estimate.

The transformation map required by a KKL observer is a solution to a specific partial differential equation (PDE), \cite{kazantzis1998, andrieu2006}. This PDE is computationally challenging to solve in practice, and finding the left inverse of the transformation map in real time is further challenging. We develop a framework for learning these maps using synthetic data generated from system dynamics and sensor measurements. 
This paper makes several key contributions to the learning-based design and implementation of KKL observers. 
We propose a learning method that employs physics-informed neural networks (PINNs) to design KKL observers and establish theoretical learning guarantees. Additionally, we provide robustness guarantees for the learned KKL observer against approximation errors and system uncertainties.
Through numerical simulations, we validate the effectiveness of our learning-based approach to KKL observer design and provide comparisons with state-of-the-art approaches.

The rest of the paper is organized as follows. Section~\ref{sec_background} provides background and related literature on KKL observers and situates our work within the existing literature.
Section~\ref{sec_prelim} reviews the theoretical foundations of KKL observers. 
Section~\ref{sec_prelim-stat-learn} reviews relevant notions in machine learning theory.
In Section~\ref{sec_learning}, we present our learning method for KKL observer design, followed by {Sections~\ref{sec_learning-guarantees} and \ref{sec_robustness}, which develop theoretical learning and robustness guarantees, respectively.} 
The effectiveness of our approach is demonstrated through numerical examples and comparisons in Section~\ref{sec_simulations}.
Finally, Section~\ref{sec_conclusion} concludes the paper with a discussion of implications and future research directions.

\textit{Notations.}
For a vector $v\in\bb R^n$, $\|v\|$ denotes its Euclidean norm and $\|v\|_\infty$ its max norm. For a matrix $M\in\bb R^{m\times n}$, $\|M\|$ denotes its spectral norm and $\cond(M)$ its condition number. For $M\in\bb R^{n\times n}$, $\eig(M)\subset\bb C$ is the set of its eigenvalues.
We denote $[p]\eqdef\{1,\dots, p\}$ for $p\in\bb Z_{>0}$.
For a signal $s:\bb R_{\geq 0}\to \bb R^n$, $s_{[0,t]}$ denotes its restriction to $[0,t]$ for $t\in\bb R_{\geq 0}$, and $\|s_{[0,t]}\|_{L_\infty}$ is its essential supremum (or $L_\infty$) norm.

\section{Background and Related Literature} \label{sec_background}

KKL observers generalize the theory of Luenberger observers \cite{luenberger1964, luenberger1966, luenberger1971} to nonlinear systems. 
Although the idea was initially proposed in \cite{shoshitaishvili1990} and \cite{shoshitaishvili1992}, KKL observers were subsequently rediscovered by Kazantis and Kravaris \cite{kazantzis1998}, who provided local guarantees around an equilibrium point of the estimation error dynamics via Lyapunov's Auxiliary Theorem. Later, \cite{krener2002b} relaxed the restrictive assumptions of \cite{kazantzis1998} to some extent; however, the analysis remained local until \cite{kreisselmeier2003} proposed the first global result under the assumption of so-called finite complexity, which also proved quite restrictive for general systems. 
Subsequently, Andrieu and Praly \cite{andrieu2006} provided a comprehensive treatment of the KKL observer problem. Their key contribution lies in relating the existence of an injective KKL transformation map via Coron's lemma to an observability-like property of the system known as backward distinguishability.
In addition, \cite{andrieu2014} proved that KKL observers converge exponentially and are tunable if the system is also differentially observable. The existence conditions are further refined in \cite{brivadis2023}, and KKL observers with contracting dynamics are developed in \cite{pachy2024}. When the observability conditions are not satisfied, \cite{bernard2024} proposes a set-valued KKL observer that estimates a set of possible state trajectories that could have produced the observed output trajectories. Finally, extensions of KKL observers to non-autonomous and controlled nonlinear systems are presented in \cite{engel2007, bernard2017, bernard2018, tran2023}; however, such systems are beyond the scope of the present paper.

The main challenge in synthesizing KKL observers for autonomous nonlinear systems is not only finding the transformation map that puts the system into a normal form, but also finding its inverse map. Both problems turn out to be challenging in practice. To this end, \cite{ramos2020, peralez2021, buisson2023, niazi2023, miao2023, tran2023, peralez2024} have proposed several methods to approximate the transformation map and its inverse via deep neural networks. 
By fixing the observer dynamics, \cite{ramos2020} proposes generating synthetic data trajectories by numerically integrating the system model and the KKL observer from random initial points. Then, using a supervised learning approach, a neural network is trained on the synthetic data to approximate the transformation map and its left inverse. For discrete-time nonlinear systems, \cite{peralez2021} proposes an unsupervised learning approach that enables proper exploration of the state space during training, whereas \cite{peralez2024} extends this approach by allowing switching observers. However, noisy measurements can lead to erroneous switching decisions, causing the observer to select the wrong observer mode or introduce instability due to chattering. Similarly, \cite{buisson2023} proposes another unsupervised learning approach to tune a KKL observer by incorporating the PDE associated with the transformation map as a design constraint. However, this unsupervised approach exhibits poor generalization capabilities when the real system's initial state deviates significantly from the conditions observed during training. When the system dynamics are partially or fully unknown, \cite{miao2023} proposes a neural ODE-based approach to design KKL observers, enabling analysis of the trade-off between convergence speed and robustness and leading to improved observer training for robust performance.

Our prior work \cite{niazi2023} leveraged physics-informed learning to design KKL observers with improved generalization and training efficiency, and \cite{cao2024} applied this approach to detect and isolate sensor faults. In contrast to other learning-based approaches, we demonstrated that our method avoids overfitting and achieves better generalization performance across the entire state space. However, \cite{niazi2023} employed a joint encoder-decoder neural network architecture to simultaneously learn the transformation map and its inverse, which leads to conflicting gradients across components of the objective function. This causes the optimization to sometimes get stuck in a bad local minimum, which degrades the observer's accuracy in some instances \cite{peralez2024}. In our current work, we introduce a sequential learning approach for the KKL observer. We first learn the transformation map and subsequently utilize it to learn its inverse. This approach improves the observer's accuracy by avoiding conflicting gradients during training.
Furthermore, we provide theoretical non-asymptotic generalization bounds for the learned maps and prove that the resulting state-estimation error is {asymptotically robust} to learning errors and system uncertainties.

\section{Preliminaries on KKL Observers} \label{sec_prelim}

\subsection{State Observation Problem}

Consider a compact set $\mc X\subset\bb R^{n_x}$ and a nonlinear system
\begin{subequations}
    \label{eq:sys}
    \begin{align}
        \dot{x}(t) &= f(x(t)), \quad x(0)=x_0 
        \label{eq:sys-state} \\
        y(t) &= h(x(t))
        \label{eq:sys-output}
    \end{align}
\end{subequations} 
where $x(t)\in\mc X$ is the state at time $t\in\bb R_{\geq 0}$, $x_0\in\mc X$ is an \textit{unknown} point from where the system is initialized, $y(t)\in\bb R^{n_y}$ is the \textit{measured} output, and $f:\mc X\to\bb R^{n_x}$ and $h:\mc X\to\bb R^{n_y}$ are smooth functions.
The state observation problem involves designing an \textit{observer}
\begin{subequations}
    \label{eq:obs-general}
    \begin{align}
        \dot{\hat z}(t) &= \Phi(\hat z(t),y(t)), \quad \hat z(0) = \hat z_0 
        \label{eq:obs-general-state} \\
        \hat x(t) &= \Psi(\hat z(t),y(t))
        \label{eq:obs-general-output}
    \end{align}
\end{subequations}
where $\hat z(t)\in\bb R^{n_z}$ is initialized at $\hat z_0\in\bb R^{n_z}$. The observer takes the output $y(t)$ of \eqref{eq:sys} as its input and provides an estimate $\hat x(t)\in\bb R^{n_x}$ of the state $x(t)$. 
Designing an observer means choosing functions $\Phi:\bb{R}^{n_z}\times\bb{R}^{n_y}\to\bb{R}^{n_z}$ and $\Psi:\bb{R}^{n_z}\times\bb{R}^{n_y}\to\bb{R}^{n_x}$ so that the estimation error 
\be
    \label{eq:est_error}
    \xi(t)\coloneq x(t)-\hat x(t)
\ee
\textit{globally} asymptotically converges to zero as $t\to\infty$. That is, for all $x_0\in\mc X$ and $\hat z_0\in\bb R^{n_z}$,
\be \label{eq:estimation-error-requirement}
\lim_{t\to\infty} \|\xi(t)\|=0.
\ee

\subsection{KKL Observer}

Designing a KKL observer involves transforming the nonlinear system~\eqref{eq:sys} to a higher-dimensional state space where its dynamics is linear up to output injection and bounded-input bounded-state stable.
This transformation $\mc T : \mc X\to\mc Z$, which must be injective\footnote{A map $\mc T:\mc X\to\bb{R}^{n_z}$ is said to be \textit{injective} if, for every $x_1,x_2\in\mc X$, $\mc T(x_1)=\mc T(x_2)$ implies $x_1=x_2$.}, maps every point $x$ in the state space $\mc X\subseteq\bb{R}^{n_x}$ of \eqref{eq:sys} to a point $z=\mc T(x)$ in the new state space $\mc Z\subseteq\bb{R}^{n_z}$, where in general $n_z\gg n_x$. The dynamics in the new state space $\mc Z$ is linear up to output injection and given by
\be \label{eq:sys_z}
    \dot{z}(t) = Az(t) + Bh(x(t)), \quad z(0) = \mc T(x_0)
\ee
where $A\in\bb{R}^{n_z\times n_z}$ and $B\in\bb{R}^{n_z\times n_y}$ are chosen such that $A$ is Hurwitz and the pair $(A,B)$ is controllable\footnote{$(A,B)$ is controllable iff $\rank\left[\arraycolsep=1pt\ba{cccc} B & AB & \dots & A^{n_z-1}B \ea\right]=n_z$.}. 
Although the transformed system \eqref{eq:sys_z} in the new coordinates is stable, the original nonlinear system~\eqref{eq:sys} need not be (Lyapunov) stable.
Moreover, \eqref{eq:sys_z} is not a linear system, but its linearity is only up to output injection, i.e., $h(x)$ is treated as an external input to \eqref{eq:sys_z}, even though $h(x)=h(\mc T^*(z))$ with $\mc T^*$ the left-inverse of $\mc T$. 

Since 
$
\dot{z}(t)=\frac{\partial \mc T}{\partial x}(x(t))\dot{x}(t)
$,
it follows from \eqref{eq:sys_z} that, for a given $A$ and $B$, $\mc T$ must satisfy the following PDE:
\be \label{eq:pde}
    \frac{\partial \mc T}{\partial x}(x(t)) f(x(t)) = A\mc T(x(t)) + Bh(x(t))
\ee
with $\mc T(0_{n_x})=0_{n_z}$.

To obtain a state estimate $\hat x(t)$ in the original coordinates $\bb{R}^{n_x}$, the map $\mc T$ must be injective, implying the existence of its left inverse $\mc T^*$, i.e., $\mc T^*(\mc T(x))=x$. When this inverse exists, the KKL observer is obtained as
\begin{subequations}
    \label{eq:kkl_observer}
    \begin{align}
        \dot{\hat z}(t) &= A\hat z(t) + By(t), \quad \hat z(0) = \hat z_0 \\
        \hat x(t) &= \mc T^*(\hat z(t)).
    \end{align}
\end{subequations}
Notice that KKL observer~\eqref{eq:kkl_observer} is a special case of \eqref{eq:obs-general}, where $\Phi(\hat z,y)$ is affine in $\hat z$ and $\Psi(\hat z,y) \eqdef \mc T^*(\hat z)$.

\subsection{Existence Results}

We explore sufficient conditions that ensure the existence of an injective transformation map $\mc T$ such that the estimate $\hat x(t)$ obtained from KKL observer~\eqref{eq:kkl_observer} satisfies the estimation error requirement \eqref{eq:estimation-error-requirement}.

Let the solution of \eqref{eq:sys-state} initialized at $x_0\in\mc X$ be $x(t;x_0)$. 

\begin{define}
    The system~\eqref{eq:sys} is \emph{forward complete} in $\mc X$ if, for every $x_0\in\mc X$, the solution $x(t;x_0)$ exists for every $t\in\bb{R}_{\geq 0}$ and remains in $\mc X$.
\end{define}

\begin{assume} \label{assumption_1}
    There exists a compact set $\mc X\subset\bb{R}^{n_x}$ such that the system~\eqref{eq:sys} is forward complete in $\mc X$.
\end{assume}

This assumption restricts our attention to nonlinear systems whose state $x(t)$ remains in a bounded domain.

\begin{define}
    A map $\mc T:\mc X\to\mc Z$ is said to be \emph{uniformly injective} if there exists a class $\mc K$ function\footnote{A function $\rho:\bb{R}_{\geq 0}\to\bb{R}_{\geq 0}$ is of class $\mc K$ if it is continuous, zero at zero, and strictly increasing. 
    It is of class $\mc K_\infty$ if it is of class $\mc K$ and unbounded.
    } 
    $\rho:\bb R_{\geq 0}\to\bb R_{\geq 0}$ such that, for every $x,\hat x\in\mc X$, 
    \be \label{eq:uniformly_injective}
    \|x-\hat x\| \leq \rho(\|\mc T(x)-\mc T(\hat x)\|).
    \ee
\end{define}

Notice that \eqref{eq:uniformly_injective} implies
$
\|z-\hat z\| \leq \varrho(\|\mc T^*(z) - \mc T^*(\hat z)\|)
$,
for every $z,\hat z\in\mc Z$ and for some class $\mc K$ function $\varrho$.

\begin{rem}
    For the existence of a KKL observer~\eqref{eq:kkl_observer} satisfying the requirement \eqref{eq:estimation-error-requirement}, it is sufficient that \eqref{eq:sys} is forward complete and the map $\mc T$ satisfying the PDE \eqref{eq:pde} is uniformly injective; see \cite[Theorem 1]{andrieu2006}.
    \hfill $\diamond$
\end{rem}

Let $\tilde{z}(t) = \hat{z}(t) - z(t)$. Since $A$ is Hurwitz, the observer error dynamics $\dot{\tilde z}(t) = A \tilde z(t)$ is globally exponentially stable in $\mathbb{R}^{n_z}$. However, the uniform injectivity of the map $\mathcal{T}$, and the corresponding existence of a class $\mathcal{K}$ function $\rho$ satisfying \eqref{eq:uniformly_injective} is guaranteed only for points $x, \hat x \in \mathcal{X}$. Because the observer is initialized with an arbitrary $\hat z_0 \in \mathbb{R}^{n_z}$, the transient trajectory $\hat{z}(t)$ does not generally evolve on the image manifold $\mathcal{T}(\mathcal{X})$. Therefore, \eqref{eq:uniformly_injective} may not hold when $\hat{z}(t) \notin \mathcal{T}(\mathcal{X})$.
Instead, convergence in the original coordinates is established asymptotically. Because the true state $x(t) \in \mathcal{X}$, its transformed counterpart $z(t) = \mathcal{T}(x(t))$ is always confined to $\mathcal{T}(\mathcal{X})$. By defining the state estimate as $\hat{x}(t) = \mathcal{T}^*(\hat{z}(t))$, where $\mathcal{T}^* : \mathbb{R}^{n_z} \to \mathbb{R}^{n_x}$ is a uniformly continuous extension of the left-inverse from $\mathcal{T}(\mathcal{X})$ to the ambient space $\mathbb{R}^{n_z}$, we guarantee asymptotic tracking. As $t \to \infty$, the distance between $\hat{z}(t)$ and the manifold $\mathcal{T}(\mathcal{X})$ vanishes. By the uniform continuity of $\mathcal{T}^*$, it follows that the estimation error $\xi(t) = x(t) - \hat{x}(t) = \mathcal{T}^*(z(t)) - \mathcal{T}^*(\hat{z}(t))$ converges to zero asymptotically as $\|z(t) - \hat{z}(t)\| \to 0$.

\begin{define}
    Given an open set $\mc O\supset\mc X$, the system~\eqref{eq:sys} is \emph{backward $\mc O$-distinguishable} in $\mc X$ if, for every pair of distinct initial conditions $x_0^1,x_0^2\in\mc X$, there exists $\tau\in\bb R_{<0}$ such that the backward solutions $x(t;x_0^1), x(t;x_0^2)\in\mc O$ exist for $t\in[\tau,0]$, and
    $
    h(x(\tau;x_0^1))\neq h(x(\tau;x_0^2)).
    $
\end{define}

Although we assumed that \eqref{eq:sys} is forward complete in $\mc X$, it may not be backward complete in $\mc X$. That is, the state trajectories of \eqref{eq:sys} may leave $\mc X$, and may even go unbounded, in negative time ($t\in\bb R_{<0}$). The notion of backward $\mc O$-distinguishability guarantees the existence of a finite negative time such that the output maps, corresponding to a pair of trajectories initialized at different points in $\mc X$, can be distinguished before any of the trajectories leaves $\mc O\supset\mc X$ in backward time. Backward $\mc O$-distinguishability is related to the notions of determinability or constructability in linear systems \cite{antsaklis2007, bernard2022}.

\begin{assume} \label{assumption_2}
    There exists an open bounded set $\mc O\supset\mc X$ such that \eqref{eq:sys} is backward $\mc O$-distinguishable in $\mc X$.
\end{assume}

Assumptions~\ref{assumption_1} and \ref{assumption_2} are sufficient for the existence of a \textit{uniformly injective} map $\mc T$ satisfying \eqref{eq:pde}. 
This result is obtained in slightly different forms in \cite{andrieu2006} and \cite{bernard2022}. The most recent result by \cite{brivadis2023} can be restated as follows.

\begin{thm}{(Brivadis et al. \cite{brivadis2023}).}
\label{thm:BOD}
    Let Assumptions~\ref{assumption_1} and \ref{assumption_2} hold. Then, for almost any $(A,B)\in\bb{R}^{n_z\times n_z} \times \bb{R}^{n_z\times n_y}$ such that $n_z=n_y(2n_x+1)$, $A$ is Hurwitz, and $(A,B)$ is controllable,
    there exists a uniformly injective map $\mc T:\mc X\to\mc Z$ that satisfies the PDE \eqref{eq:pde}.
\end{thm}

This theorem guarantees the existence of a uniformly injective map $\mc T$ and its left-inverse $\mc T^*$ such that the estimate $\hat x(t)$ obtained from the KKL observer~\eqref{eq:kkl_observer} asymptotically converges to the true state $x(t)$. Therefore, relying on Assumptions~\ref{assumption_1} and \ref{assumption_2}, and fixing $n_z = n_y(2n_x+1)$, $A\in\bb{R}^{n_z\times n_z}$ a Hurwitz matrix, and $B\in\bb{R}^{n_z\times n_y}$ such that $(A,B)$ is controllable, we can learn the transformation map $\mc T$ and its left-inverse $\mc T^*$.

\section{Preliminaries on Learning Theory}
\label{sec_prelim-stat-learn}
We revisit the standard framework of statistical learning theory that will be relevant to this paper.

\subsection{Learning Problem}

Consider a supervised learning problem where the goal is to learn a mapping $\hat{\mc T}: \mc X \to \mc Z$ from an input space $\mc X$ to an output space $\mc Z$. We assume an unknown, fixed probability distribution $\mathcal{D}$ over $\mc X \times \mc Z$. The quality of a prediction $\hat{\mc T}(x)$ for a true output $z$ is measured by a loss function $L(\hat{\mc T}(x), z)$. 
The goal is to find a function $\hat{\mc T}$ that minimizes the \textit{true} risk
$$
R(\hat{\mc T}) = \mathbb{E}_{(x,z) \sim \mathcal{D}} [L(\hat{\mc T}(x), z)]
$$
which is the expected loss over the true distribution $\mathcal{D}$.
The minimal possible risk is the Bayes risk, $R^* = \inf_{\hat{\mc T}} R(\hat{\mc T})$, where the infimum is over all measurable functions \cite{bach2024}.
Since $\mathcal{D}$ is unknown, we must learn from a finite training sample $S = \{(x_i, z_i)\}_{i=1}^N$ of $N$ examples drawn independently and identically distributed (i.i.d.) from $\mathcal{D}$. On this sample $S$, a learning algorithm is supposed to minimize the empirical risk
$$
\hat R_S(\hat{\mc T}) = \frac{1}{N} \sum_{i=1}^N L(\hat{\mc T}(x_i), z_i).
$$

\subsection{Hypothesis Class and Error Decomposition}

A learning algorithm searches for a solution within a predefined hypothesis class $\mathcal{H}$, which is a set of functions $\hat{\mc T}: \mc X \to \mc Z$. For example, $\mathcal{H}$ can be the set of all functions representable by a specific neural network architecture.
The empirical risk minimizer is the hypothesis $\hat{\mc T}_S \in \mathcal{H}$ that minimizes the empirical risk, i.e., 
$
\hat{\mc T}_S = \arg \min_{\hat{\mc T} \in \mathcal{H}} \hat R_S(\hat{\mc T}).
$
The \emph{excess risk} $R(\hat{\mc T}_S) - R^*$ of the learned hypothesis, where $R^*$ is the Bayes risk, can be decomposed into two fundamental components: approximation error $\inf_{\hat{\mc T} \in \mathcal{H}} R(\hat{\mc T}) - R^*$ and estimation error $R(\hat{\mc T}_S) - \inf_{\hat{\mc T} \in \mathcal{H}} R(\hat{\mc T})$.
The approximation error is deterministic and measures how well the chosen class $\mathcal{H}$ can, in principle, approximate the true optimal (Bayes) predictor. It is independent of the data size $N$. On the other hand, the estimation error is random (it depends on the sample $S$) and measures the penalty for having only a finite sample of $N$ data points instead of the true distribution $\mathcal{D}$. A central goal of learning theory is to bound the estimation error, which requires a measure of the richness (or complexity) of the hypothesis class $\mathcal{H}$.

\subsection{Pseudo-Dimension and Generalization Bounds}

For real-valued functions, such as those in regression or in our KKL observer learning problem, the VC-dimension \cite{vapnik1971} (used for binary classification) is extended to the pseudo-dimension \cite{pollard1990, haussler1992}. This concept is defined not on the hypothesis class $\mathcal{H}$ itself, but on the associated loss class $\mathcal{G} = \{ (x, z) \mapsto L(\hat{\mc T}(x), z) : \hat{\mc T} \in \mathcal{H} \}$.

\begin{define}{(Shattering and Pseudo-dimension \cite{mohri2018}).}
    A set of $d$ points $\{(x_i, z_i)\}_{i=1}^d$ is \emph{shattered} by the loss class $\mathcal{G}$ if there exist thresholds $\tau_1, \dots, \tau_d \in \mathbb{R}$ such that for all $2^d$ possible binary labelings $b \in \{-1, +1\}^d$, there exists $\mc T \in \mathcal{H}$ whose corresponding loss $g \in \mathcal{G}$ satisfies
    $$
    \forall i \in \{1, \dots, d\}, \quad \mathrm{sign}(g(x_i, z_i) - \tau_i) = b_i.
    $$
    The \emph{pseudo-dimension} of $\mathcal{G}$, denoted $\mathrm{Pdim}(\mathcal{G})$, is the size $d$ of the largest set that can be shattered by $\mathcal{G}$.
\end{define}

The pseudo-dimension provides a combinatorial measure of the complexity of $\mathcal{H}$ via its loss class $\mathcal{G}$. A finite pseudo-dimension $d$ ensures that the estimation error will decrease as the sample size $N$ increases. This is formalized in a generalization bound below.
\begin{lem} \label{lem:gen-bound-mohri}
    {(Generalization Bound for Bounded Regression \cite[Theorem 11.8]{mohri2018}).}
    Let $\mathcal{G}$ be a loss class with pseudo-dimension $d = \mathrm{Pdim}(\mathcal{G}) < \infty$. Assume the loss $L$ is bounded by a constant $M$. Then, for any $\delta > 0$, with probability at least $1-\delta$ over the draw of an i.i.d. sample $S$ of size $N$, it holds that for every $\hat{\mc T} \in \mathcal{H}$,
    $$
    R(\hat{\mc T}) \le \hat R_S(\hat{\mc T}) + M\sqrt{\frac{2d \ln(eN/d)}{N}} + M\sqrt{\frac{\ln(1/\delta)}{2N}}.
    $$
\end{lem}

This lemma provides an explicit, non-asymptotic bound on the estimation error. The complexity term 
$$
\mc C(\mathcal{H}, N, \delta) \triangleq M\sqrt{\frac{2d \ln(eN/d)}{N}} + M\sqrt{\frac{\ln(1/\delta)}{2N}}
$$
depends on the hypothesis class $\mathcal{H}$ via $d$ and the sample size $N$. For the learning procedure of KKL observers developed in this paper, we use Lemma~\ref{lem:gen-bound-mohri} to derive explicit non-asymptotic error bounds. 

\section{Learning Method for KKL Observers} \label{sec_learning}

Designing KKL observers involves (a)~finding the injective map $\mc T: \mc X\to \mc Z$ that satisfies PDE \eqref{eq:pde}, so that the nonlinear system~\eqref{eq:sys} admits a linear up to output injection representation \eqref{eq:sys_z}, and (b)~finding its left-inverse $\mc T^*$, so that a state estimate can be obtained in the original state coordinates. For finding $\mc T$, one must solve PDE \eqref{eq:pde}, whose explicit solution derived in \cite{andrieu2006} is given by
\be \label{eq:T_solution}
\mc T(x) = \int_{-\infty}^0 \exp(A\tau) Bh(\breve{x}(\tau;x)) \mathrm{d}\tau
\ee
where $\breve{x}(\tau;x)\in\mc X$ is the backward solution trajectory initialized at $x\in\mc X$, for $\tau\in\bb R_{\leq 0}$, to the modified dynamics $\dot{\breve{x}}(\tau)=g(\breve{x}(\tau))$ with $g(\breve{x}(\tau))=f(\breve{x}(\tau))$ if $\breve{x}(\tau)\in\mc X$ and $g(\breve{x}(\tau))=0$ otherwise. 
However, as pointed out by \cite{bernard2022}, computing \eqref{eq:T_solution} is not practically possible due to the inaccessibility of the backward output map $h(\breve{x}(\tau;x))$ for $\tau<0$ and the infeasibility of computing the integral \eqref{eq:T_solution} for every initial point $x\in\mc X$.
Note that, even if $\mc T$ is known in some other form\footnote{See \cite{bernard2018} for some examples.} than \eqref{eq:T_solution}, finding the left-inverse $\mc T^*$ is difficult both analytically and numerically \cite{andrieu2021}.
To overcome these challenges, we propose learning the maps $\mc T$ and $\mc T^*$.

\subsection{Learning Problem}
We introduce two expectation operators corresponding to the underlying data distributions. Let $\mathcal{D}_{\mathcal{X}}$ be the true data distribution over the compact state space $\mathcal{X}$. The first expectation operator is defined as
$$
\mathbb{E}_{x \sim \mathcal{D}_{\mathcal{X}}}[\cdot] = \int_{\mathcal{X}} (\cdot) \ \mathrm{d}\mathcal{D}_{\mathcal{X}}(x).
$$
The injective map $\mathcal{T}: \mathcal{X} \to \mathcal{Z}$ induces a pushforward measure $\mu_{\mathcal{T}}$ on the observer space $\mathcal{Z}$, defined by $\mu_{\mathcal{T}}(Z) = \mathcal{D}_{\mathcal{X}}(\mathcal{T}^{-1}(Z))$ for any measurable set $Z \subset \mathcal{Z}$, {where $\mathcal{T}^{-1}(Z) = \{ x \in \mathcal{X} \mid \mathcal{T}(x) \in Z \}$ is the set-theoretic preimage}. 
The second expectation operator is defined with respect to this measure as
$$
\mathbb{E}_{z \sim \mu_{\mathcal{T}}}[\cdot] = \int_{\mathcal{Z}} (\cdot) \ \mathrm{d}\mu_{\mathcal{T}}(z).
$$
Now, we define the risk and error functions associated with the learning of a KKL observer.

Our goal is to find approximations $\hat{\mathcal{T}}_{\theta} \in \mathcal{H}_{\theta}$ and $\hat{\mathcal{T}}_{\eta}^* \in \mathcal{H}_{\eta}$ for $\mc T$ and $\mc T^*$, where $\mathcal{H}_{\theta}$ and $\mathcal{H}_{\eta}$ are the hypothesis classes (e.g., neural network architectures) for the forward and inverse maps. 
The learned forward map $\hat{\mathcal{T}}_{\theta}$ must satisfy two conditions: (i)~it must map $x$ to the correct $z = \mathcal{T}(x)$, and (ii)~it must satisfy the governing PDE \eqref{eq:pde}. 
The \textit{forward map error} is given by
\be \label{eq:forward-error-RT}
    R_{\mc T}(\theta) = \bb E_{x\sim \mc D_{\mc X}} \big[
    \| \mc T(x) - \hat{\mc T}_\theta(x) \|^2 \big]
\ee
and the PDE error by
$
    R_\mathrm{PDE}(\theta) = \bb E_{x\sim \mc D_{\mc X}} \big[ \| \mc P_\theta(x) \|^2 \big]
$
where $\mathcal{P}_{\theta}(x)$ is the PDE residual
\begin{equation}
    \label{eq:pde_residual}
    \mc P_\theta(x) \coloneq \frac{\partial \hat{\mc T}_\theta}{\partial x}(x) f(x) -A\hat{\mc T}_\theta(x) -Bh(x).
\end{equation}
We combine these into the \textit{physics-informed risk}
\begin{equation}
\label{eq:true-risk-R1-theta}
R_1(\theta) = R_{\mc T}(\theta) + \nu R_\mathrm{PDE}(\theta)
\end{equation}
where $\nu > 0$ is a hyperparameter.

The learned inverse map $\hat{\mc T}_\eta^*$ must approximate the true inverse $\mc T^*$ by minimizing the \textit{inverse map error}
\be \label{eq:inv-error-RT*}
    R_{\mc T^*}(\eta) = \bb E_{z\sim \mu_{\mc T}} \big[ \| \mc T^*(z) - \hat{\mc T}_\eta^*(z) \|^2 \big]
\ee
where $\mu_{\mc T}$ is the pushforward measure on the observer space $\mathcal{Z}$.
{ However, minimizing the true inverse error $R_{\mathcal{T}^{*}}(\eta)$ is infeasible because the true inverse $\mathcal{T}^*$ and the measure $\mu_{\mathcal{T}}$ are unknown. To overcome this, we train the inverse map to minimize a computable surrogate. Specifically, given the forward map $\hat{\mathcal{T}}_{\theta}$, the learned inverse $\hat{\mathcal{T}}_{\eta}^{*}$ must reconstruct the state $x$ from the estimated latent state $\hat{\mathcal{T}}_{\theta}(x)$ by minimizing the \textit{reconstruction risk}}
\begin{equation}
\label{eq:true-risk-R2-eta}
    R_2(\eta \mid \theta) = \mathbb{E}_{x \sim \mathcal{D}_{\mathcal{X}}} \big[ \|x - \hat{\mathcal{T}}_{\eta}^*(\hat{\mathcal{T}}_{\theta}(x))\|^2 \big].
\end{equation}
As we will establish later in Lemma~\ref{lem:R-eta-bound}, minimizing $R_2(\eta\mid \theta)$ provides an upper bound on the true inverse map error $R_{\mathcal{T}^{*}}(\eta)$, thereby justifying its use as the training objective.
Our goal is to find parameters $\theta$ and $\eta$ that sequentially minimize $R_1(\theta)$ and $R_2(\eta \mid \theta)$, respectively. Since the true data distribution $\mathcal{D}_{\mathcal{X}}$ and the true maps $\mathcal{T}$ and $\mc T^*$ are unknown, we minimize empirical approximations of these risks using synthetically generated data.

\subsection{Data Generation Procedure} \label{subsec_datagen}

Before defining empirical approximations of these risks, we describe a heuristic for generating synthetic data.
Since the observer matrices $A$ and $B$ are assumed to be fixed and the system functions $f$ and $h$ are known, it is possible to generate trajectories of $x$ and $z$ numerically. 
However, given $x_0\in\mc X$, one cannot guess the initial condition $z(0) = {\mc T} (x_0)$ because $\mc T$ is unknown.
Following \cite{ramos2020}, one could mitigate this because, for an arbitrary $z(0)=z_0$, the solution of \eqref{eq:sys_z} satisfies
\[
    \|z(t;z_0)\| \leq \|\exp(At)z_0\| + \int_0^t \| \exp(A(t-\tau)) By(\tau) \| \mathrm{d}\tau.
\]
Since $A$ is Hurwitz, $\|\exp(At)z_0\| \to 0$ exponentially fast, and the effect of the initial condition $z_0$ vanishes from $z(t;z_0)$ over time $t$. We have $\left\| \exp(At)z_0 \right\| \leq \cond(V)e^{\lambda_{\min}(A)t} \| z_0 \|$, where $V$ is obtained from the eigendecomposition of $A=V\Lambda V^{-1}$ and $\lambda_{\min}(A)\!\in\!\min_{\lambda\in\eig(A)}|\text{Re}(\lambda)|$.
For any $\varepsilon>0$, there exists time $t_*$ such that $\left\| \exp(At)z_0 \right\|\leq \varepsilon$, for all $t \ge t_*$, where
$
    t_* \eqdef t_*(\varepsilon, z_0) = \frac{1}{\lambda_{\min}(A)} \ln\left(\frac{\varepsilon}{\cond(V) \| z_0 \|}\right).
$
Therefore,
\[
    \forall t \geq t_*, \quad \|z(t;z_0)\| \leq \varepsilon + \int_0^t \| \exp(A(t-\tau)) By(\tau) \| \mathrm{d}\tau.
\]
As time progresses, the trajectory $z(t;z_0^i)$ becomes almost independent of the initial condition $z_0$. This observation leads to the following data generation procedure.

\textit{Initial condition sampling:} We sample $p\in\bb Z_{>0}$ initial points $\left\{ (x_0^i,z_0^i) \right\}_{i \in [p]}$ uniformly in $\mc X \times \mc Z$.

\textit{Trajectory generation:} For all $i \in [p]$, we simulate \eqref{eq:sys} and \eqref{eq:sys_z} with initial condition $(x_0^i, z_0^i)$ from $t_1 = 0$ to $t_{\tau} = T$, where $T\in\bb R_{>0}$ is chosen to be 
large.

\textit{Truncation:} For all $i \in [p]$, the samples of the $z$-trajectory for $t_k \geq t_{k_*}$ are kept, and the rest of the samples are discarded, where
\[
    k_* \eqdef \min \left\{ k \ | \ t_k \geq \max_{i\in[p]} t_*(\varepsilon, z_0^i) \right\}.
\]

This procedure yields the training set 
$$
S_\mathrm{data} = \{(x(t_k; x_0^i), z(t_k; z_0^i))\}_{i \in [p], k \in \{k_*, \dots, \tau\}}
$$ 
containing truncated trajectories with $N_\mathrm{data} = p (\tau - k_* + 1)$ data pairs. We further sample $q\in \mathbb Z_{>0}$ points $x_0^j$ {independently} and generate trajectories
$$
S_\mathrm{pde} = \{x(t_k; x_0^j)\}_{j\in [q], k\in\{1,\dots,\tau\}}
$$ 
with $N_\mathrm{pde} = q \tau$ collocation points for the PDE. The data $S_1=\{S_\mathrm{data}, S_\mathrm{pde}\}$ is used for learning the forward map $\hat{\mc T}_\theta$ with total number of data points $N_1 = N_\mathrm{data} + N_\mathrm{pde}$.
After training the forward map $\hat{\mathcal{T}}_{\theta}$, its parameters $\theta$ are fixed. We then generate a new dataset $S_2 = \{(z'_j, x'_j)\}_{j=1}^{N_2}$ of size $N_2$, where $x'$ are randomly sampled points in $\mc X$, and the corresponding features $z'_j = \hat{\mc T}_\theta(x'_j)$ are generated by fixing the learned forward map.

\begin{figure}[!ht]
    \centering
    \includegraphics[width=0.7\linewidth]{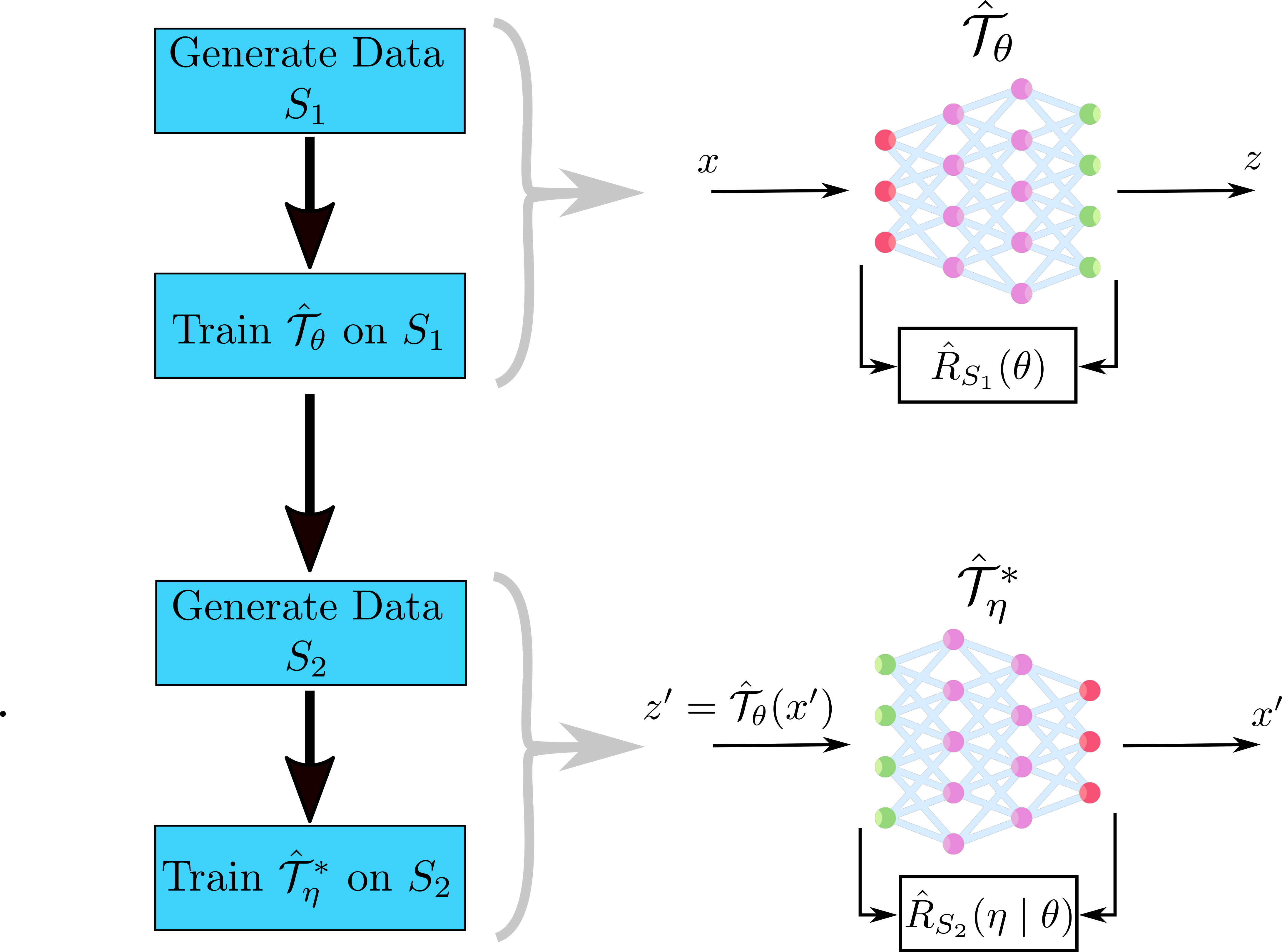}
    \caption{{Overview of the proposed KKL observer learning algorithm.}}
    \label{fig:workflow}
\end{figure}

\subsection{Empirical Risk Minimization} 
\label{subsec_emp-risk-min}

We approximate the true risks $R_1$ and $R_2$ in \eqref{eq:true-risk-R1-theta} and \eqref{eq:true-risk-R2-eta} with their empirical counterparts, computed from finite datasets $S_1$ and $S_2$. First, we minimize empirical physics-informed risk $\hat R_{S_1}$ using $S_1 = \{S_\mathrm{data}, S_\mathrm{pde}\}$ with 
\begin{equation}
\label{eq:emp-risk-theta}
    \hat R_{S_1}(\theta) = \frac{1}{N_\mathrm{data}} \sum_{(x,z) \in S_\mathrm{data}} \|z - \hat{\mathcal{T}}_{\theta}(x)\|^2
    + \frac{\nu}{N_\mathrm{pde}} \sum_{x \in S_\mathrm{pde}} \|\mathcal{P}_{\theta}(x)\|^2
\end{equation}
where $\mc P_\theta(x)$ is defined in \eqref{eq:pde_residual}.
This risk is the computable objective function for the {forward} map and approximates the true risk in \eqref{eq:true-risk-R1-theta}.
Thus, the first-stage empirical risk minimization problem is formulated as 
\be \label{eq:hat-theta-star}
\min_{\theta \in \mathcal{H}_{\theta}} \hat R_{S_1}(\theta).
\ee

We follow the sequential approach \cite{peralez2021}. Fixing the trained forward map $\hat{\mathcal{T}}_{\theta}$, we generate a new dataset $S_2 = \{(z'_j, x'_j)\}_{j=1}^{N_2}$, where $z'_j = \hat{\mc T}_\theta(x'_j)$, and minimize the empirical reconstruction risk
\begin{equation}
\label{eq:emp-risk-eta}
    \hat R_{S_2}(\eta \mid \theta) = \frac{1}{N_2} \sum_{j=1}^{N_2} \|x'_j - \hat{\mathcal{T}}_{\eta}^*(z'_j)\|^2
\end{equation}
which is the computable objective function for the inverse map and approximates the true risk in \eqref{eq:true-risk-R2-eta}. Thus, the second-stage empirical risk minimization problem is
\be \label{eq:hat-eta-star}
\min_{\eta \in \mathcal{H}_{\eta}} \hat R_{S_2}(\eta \mid \theta).
\ee

\subsection{Remarks on the Learning Method}

Fig.~\ref{fig:workflow} illustrates the workflow of proposed learning algorithm. Our procedure deconstructs the autoencoder framework used in previous works, such as \cite{niazi2023}, into sequentially learning the maps $\hat{\mc T}_\theta$ and $\hat{\mc T}_\eta^*$. This idea serves to address two fundamental issues: 

\begin{enumerate}
    \item The joint encoder-decoder structure is inherently susceptible to conflicting gradients between components of the objective function, which might result in the optimization getting stuck in bad local minima. This degrades the accuracy of the learned solutions and creates the risk of failing to enforce the latent output of the autoencoder to reside within $\mathcal{Z}$, leading to a discrepancy between the input distributions of $\hat{\mathcal{T}}_\eta^*$ during operation.
    
    \item The data generation process is dependent on the truncation of the simulated trajectories to get rid of the transient effects due to the initialization of $z$-trajectories using the contraction property of \eqref{eq:sys_z}. However, this procedure does not allow us to freely choose initial condition pairs $(x_0, z_0)$, which might result in uneven sampling of certain regions of the state space. This issue has previously been addressed in \cite{buisson2023} and \cite{niazi2023} by first simulating the system backward in time from the chosen initial conditions in $\mathcal{X}$, then obtaining the corresponding values in $\mathcal{Z}$ by simulating forward from negative time to $t = 0$. However, this setup is inherently infeasible for stable systems, and some unstable systems as well, that blow up in negative time.
\end{enumerate}

In our sequential procedure, we first learn $\theta$ for the forward map $\hat{\mathcal{T}}_\theta$ by solving \eqref{eq:hat-theta-star} on the dataset $S_1$. The learned forward map is then fixed to generate a new dataset $S_2$.
The parameters $\eta$ for the inverse map $\hat{\mathcal{T}}_\eta^*$ are then learned by solving \eqref{eq:hat-eta-star} on $S_2$. 
Decoupling the training of $\hat{\mathcal{T}}_\theta$ and $\hat{\mathcal{T}}_\eta^*$ provides a significant practical advantage over joint training strategies. 
In joint training, the input distribution $z'=\hat{\mathcal{T}}_\theta(x')$ provided to the inverse network constantly shifts as $\theta$ is updated, creating a moving target. 
In a sequential approach, we first fix $\hat{\mathcal{T}}_\theta$ and subsequently train $\hat{\mathcal{T}}^*_\eta$ to act as the left-inverse of this learned approximation, rather than the true map $\mathcal{T}$. Assuming $\hat{\mathcal{T}}_\theta$ closely satisfies the KKL PDE, this guarantees that the stationary input distribution $z'$ seen by $\hat{\mathcal{T}}^*_\eta$ during training accurately reflects the distribution of latent states it will process during online inference. While minimizing the empirical reconstruction risk \eqref{eq:emp-risk-eta} targets a surrogate map, this does not compromise the validity of the approach; Lemmas~\ref{lem:R-eta-bound} and \ref{lem:R-rec-bound} in the next section establish that minimizing this surrogate risk bounds the true inverse-map error $R_{\mathcal{T}^*}$.

The formulation of the empirical physics-informed risk $\hat R_{S_1}(\theta)$ in \eqref{eq:emp-risk-theta} constitutes a semi-supervised learning framework. The data-fit term $\|z - \hat{\mathcal{T}}_{\theta}(x)\|^2$ is computed on the labeled data $S_\mathrm{data}$. 
Conversely, the physics-informed risk term $\|\mathcal{P}_{\theta}(x)\|^2$ is evaluated on the unlabeled collocation data $S_\mathrm{pde}$. 
This semi-supervised structure allows us to enforce the physical constraint \eqref{eq:pde} over a large, densely-sampled portion of the state space ($N_\mathrm{pde} \gg N_\mathrm{data}$), which acts as a powerful data-agnostic regularizer and prevents overfitting to the labeled $S_\mathrm{data}$.

\section{Non-Asymptotic Learning Guarantees}
\label{sec_learning-guarantees}
The learning method proposed in Section~\ref{sec_learning} computes the approximations $\hat{\mathcal{T}}_{\theta}$ and $\hat{\mathcal{T}}_{\eta}^*$. A critical question remains: does this learning procedure provide any guarantee that the resulting maps are ``good'' approximations of the true, unknown maps $\mathcal{T}$ and $\mathcal{T}^*$? The practical utility of the learned observer depends on this approximation quality. If the learned inverse map $\hat{\mathcal{T}}_{\eta}^*$ is a poor approximation of the true $\mathcal{T}^*$, the resulting estimate $\hat{x}(t) = \hat{\mathcal{T}}_{\eta}^*(\hat{z}(t))$ will be inaccurate, irrespective of the observer's dynamic properties.

This section establishes guarantees for our learning-based design of KKL observers. We show that the approximation error of the learning procedure is bounded. 
This error is quantified by the inverse map error $R_{\mc T^*}(\eta)$ defined in \eqref{eq:inv-error-RT*}.
Proving that $R_{\mc T^*}$ can be made small provides the foundation for our learning method.

\subsection{Lipschitz Assumption}
Given that the neural network's activation functions are Lipschitz continuous, the neural network map is also Lipschitz continuous. Thus, it is natural to make the following assumption.

\begin{assume}
    \label{assume-Lipschitz}
    The neural network $\hat{\mc T}_\eta^*$ is Lipschitz continuous over $\mc Z$, where $\mc Z\supseteq \mc T(\mc X)$. In particular, for every $z,\hat z\in\mc Z$, there exists $\ell_\eta\in\bb{R}_{>0}$ such that
    \be \label{eq:inverse_T_lipschitz}
    \|\hat{\mc T}_\eta^*(z)-\hat{\mc T}_\eta^*(\hat z)\| \leq \ell_\eta \|z-\hat z\|.
    \ee
\end{assume}

Because $\hat{\mathcal{T}}_\eta^*$ is parameterized as a neural network (e.g., composed of dense layers with bounded weights), this Lipschitz property extends globally over the entire ambient space $\mathbb{R}^{n_z}\supseteq\mathcal Z$. As we will see in Section~\ref{sec_robustness}, this property is critical: it ensures that our mathematical bounds remain valid even when transient estimation errors drive the observer state $\hat{z}(t)$ far outside the image manifold $\mathcal{T}(\mathcal{X})$.
Estimating a neural network's Lipschitz constant is a topic of interest in the machine learning community \cite{szegedy2013, virmaux2018, fazlyab2019, jordan2020}. For feedforward ReLU networks, \cite{ebihara2024} shows that the local Lipschitz constant estimation problem can be reduced to a semidefinite program.

\subsection{Generalization Guarantees}
Recall that our learning method is a sequential procedure. First, we learn $\hat{\mathcal{T}}_{\theta}$ by minimizing the empirical physics-informed risk $\hat R_{S_1}(\theta)$. Then, we learn $\hat{\mathcal{T}}_{\eta}^*$ by minimizing the empirical reconstruction risk $\hat R_{S_2}(\eta \mid \theta)$. 
The inverse map error $R_{\mc T^*}$ is thus affected by the outcomes of both stages. We relate $R_{\mc T^*}$ to the reconstruction risk $R_2(\eta \mid \theta)$ in \eqref{eq:true-risk-R2-eta}, which is empirically minimized in the second stage, and the forward map error $R_{\mc T}(\theta)$ in \eqref{eq:forward-error-RT}, which is propagated from the first stage.

\begin{lem}
    \label{lem:R-eta-bound}
    Let Assumption~\ref{assume-Lipschitz} hold. Then, the true inverse error $R_{\mc T^*}(\eta)$ in \eqref{eq:inv-error-RT*} is bounded by
    \begin{equation}
        \label{eq:inv-error-bound}
        R_{\mc T^*}(\eta) \le 2 R_2(\eta \mid \theta) + 2 \ell_{\eta}^2 R_{\mc T}(\theta)
    \end{equation}
    where $R_2(\eta \mid \theta)$ is the reconstruction risk \eqref{eq:true-risk-R2-eta}, $R_{\mc T}(\theta)$ is the forward map error \eqref{eq:forward-error-RT}, and $\ell_\eta$ is the Lipschitz constant in \eqref{eq:inverse_T_lipschitz}.
\end{lem}
\begin{proof}
    Let $\| \cdot \|_{L_2(\mathcal{D}_{\mathcal{X}})}$ denote the $L_2$ norm with respect to the data distribution $\mathcal{D}_{\mathcal{X}}$. Then,
    \begin{align*} 
        \sqrt{R_{\mc T^*}} &= \| \mathcal{T}^*(\mathcal{T}) - \hat{\mathcal{T}}_{\eta}^*(\mathcal{T}) \|_{L_2(\mathcal{D}_{\mathcal{X}})} \\ 
        &\le \| \mathcal{T}^*(\mathcal{T}) - \hat{\mathcal{T}}_{\eta}^*(\hat{\mathcal{T}}_{\theta}) \|_{L_2(\mathcal{D}_{\mathcal{X}})} + \| \hat{\mathcal{T}}_{\eta}^*(\hat{\mathcal{T}}_{\theta}) - \hat{\mathcal{T}}_{\eta}^*(\mathcal{T}) \|_{L_2(\mathcal{D}_{\mathcal{X}})} \\
        &\le \sqrt{R_2(\eta \mid \theta)} + \ell_{\eta} \|(\hat{\mathcal{T}}_{\theta} - \mathcal{T}) \|_{L_2(\mathcal{D}_{\mathcal{X}})} \\
        &= \sqrt{R_2(\eta \mid \theta)} + \ell_{\eta} \sqrt{R_{\mc T}} 
        \end{align*}
        where we used $x = \mc T^*(\mc T(x))$, the triangle inequality, Assumption~\ref{assume-Lipschitz}, and definitions \eqref{eq:true-risk-R2-eta} and \eqref{eq:forward-error-RT}.
        Squaring both sides and using the inequality $(a+b)^2 \le 2a^2 + 2b^2$ gives the desired result \eqref{eq:inv-error-bound}. 
\end{proof}

Lemma~\ref{lem:R-eta-bound} states that the inverse map error $R_{\mc T^*}$ is bounded by the reconstruction risk $R_2(\eta \mid \theta)$ and the forward error $R_{\mc T}$, amplified by the Lipschitz constant of the learned inverse map. We now bound these two terms using the generalization bound presented in Section~\ref{sec_prelim-stat-learn}.

\begin{lem}
    \label{lem:R-rec-bound}
    Let $\mathcal{H}_{\eta}$ be the hypothesis class for the learned inverse map $\hat{\mc T}_\eta^*$. Let $\mathcal{G}_{\eta} = \{ (x', z') \mapsto \|x' - \hat{\mathcal{T}}_{\eta}^*(z')\|^2 : \hat{\mathcal{T}}_{\eta}^* \in \mathcal{H}_{\eta} \}$ be the associated loss class, with finite pseudo-dimension $d_{\eta} = \mathrm{Pdim}(\mathcal{G}_{\eta})$, where $z'=\hat{\mc T}_\theta(x)$. Let $N_2$ be the number of data points in the dataset $S_2=\{(x_j',z_j')\}$. Assume the empirical reconstruction risk $\hat R_{S_2}(\eta \mid \theta)$ is bounded by $M_{\eta}$. Then, for any $\delta > 0$, with probability at least $1-\delta$, we have
    \begin{equation}
        \label{eq:recon-error-bound}
        R_2({\eta}\mid \theta) \le \hat{R}_{S_2}({\eta}\mid \theta) + \mathcal{C}_\eta(N_2, \mathcal{H}_\eta, \delta)
    \end{equation}
    where $\mc C_{\eta}(N_2, \mathcal{H}_{\eta}, \delta) = M_{\eta}\sqrt{\frac{2d_{\eta} \ln(eN_2/d_{\eta})}{N_2}} + M_{\eta}\sqrt{\frac{\ln(1/\delta)}{2N_2}}$, $R_2$ defined in \eqref{eq:true-risk-R2-eta}, and $\hat R_{S_2}$ in \eqref{eq:emp-risk-eta}.
\end{lem}

\begin{lem}
    \label{lem:R-pinn-bound}
    Let $\mathcal{H}_{\theta}$ be the hypothesis class for the learned forward map $\hat{\mc T}_\theta$. Let $\mathcal{G}_{\theta}$ be the associated loss class for the empirical physics-informed risk $\hat R_{S_1}(\theta)$ in \eqref{eq:emp-risk-theta}, with finite pseudo-dimension $d_{\theta} = \mathrm{Pdim}(\mathcal{G}_{\theta})$. Let $N_1$ be the number of data points in $S_1$. Assume $\hat R_{S_1}(\theta) \leq M_{\theta}$. Then, for any $\delta > 0$, with probability at least $1-\delta$, 
    \begin{equation}
        \label{eq:R-pinn-bound}
        R_1(\theta) \le \hat{R}_{S_1}(\theta) + \mathcal{C}_\theta(N_1, \mathcal{H}_\theta, \delta)
    \end{equation}
    where $\mc C_{\theta}(N_1,\mc H_\theta, \delta) = M_{\theta}\sqrt{\frac{2d_{\theta} \ln(eN_1/d_{\theta})}{N_1}} + M_{\theta}\sqrt{\frac{\ln(1/\delta)}{2N_1}}$, $R_1$ defined in \eqref{eq:true-risk-R1-theta}, and $\hat R_{S_1}$ in \eqref{eq:emp-risk-theta}. 
    
    Moreover, since $R_{\mc T}(\theta) \le R_1(\theta)$, the right-hand side of \eqref{eq:R-pinn-bound} also provides a high-probability bound on $R_{\mc T}$.
\end{lem}

Lemmas~\ref{lem:R-rec-bound} and \ref{lem:R-pinn-bound} are direct applications of the generalization bound in Lemma~\ref{lem:gen-bound-mohri} to the empirical risk minimization problems \eqref{eq:hat-eta-star} and \eqref{eq:hat-theta-star}, respectively.
In these lemmas, applying the generalization bounds requires the empirical risks to be bounded by constants $M_\theta$ and $M_\eta$ uniformly across the hypothesis classes $\mathcal{H}_\theta$ and $\mathcal{H}_\eta$. While the state space $\mathcal{X}$ is compact (Assumption~\ref{assumption_1}), leaving the neural network parameter space unconstrained would allow the supremum of the loss to diverge. Therefore, we explicitly restrict $\mathcal{H}_\theta$ and $\mathcal{H}_\eta$ to classes of neural networks whose parameters are bounded in norm. This bounded-parameter assumption is standard in statistical learning theory to guarantee finite loss bounds \cite[Ch. 11]{mohri2018}, and it is the foundational mechanism for controlling the Rademacher complexity of neural networks \cite[Ch. 9]{bach2024}. In practice, this aligns with standard optimization techniques such as $\ell_2$ regularization \cite[Ch. 7]{goodfellow2016}. Coupled with the continuity of the system maps and activation functions over the compact set $\mathcal{X}$, this bounded-parameter assumption formally guarantees the existence of finite global upper bounds $M_\theta$ and $M_\eta$.

\subsection{Main Theorem and Discussion}

We provide a high-probability bound on the true $L_2$ approximation error $R_{\mc T^*}$ of the learned inverse map $\hat{\mathcal{T}}_\eta^*$ by combining the results of the above lemmas.
\begin{thm}
    \label{thm:R-eta-bound}
    Let $\hat{\mathcal{T}}_{\theta} \in \mathcal{H}_{\theta}$ and $\hat{\mathcal{T}}_\eta^* \in \mathcal{H}_{\eta}$ be the maps learned by the proposed learning method. Let Assumption~\ref{assume-Lipschitz} hold.
    Then, for any $\delta>0$, with probability at least $1-2\delta$, it holds that
    \begin{equation}
        \label{eq:inv-error-RT*-bound-2}
        R_{\mathcal{T}^*}(\eta) \le 2\left(\hat{R}_{S_2}(\eta\mid \theta) + \mathcal{C}_\eta(N_2, \mathcal{H}_\eta, \delta)\right) 
        + 2\ell_\eta^2\left(\hat{R}_{S_1}(\theta) + \mathcal{C}_\theta(N_1, \mathcal{H}_\theta, \delta)\right).
    \end{equation}
\end{thm}

The proof follows from Lemmas~\ref{lem:R-eta-bound}, \ref{lem:R-rec-bound}, and \ref{lem:R-pinn-bound}. Below, we discuss some implications of the theorem and present guidelines for the learning algorithm. 

Theorem~\ref{thm:R-eta-bound} provides a non-asymptotic bound on the inverse map error $R_{\mathcal T^*}$. It shows that $R_{\mathcal T^*}$ can be made small, provided the empirical approximation errors, $\hat R_{S_1}$ and $\hat R_{S_2}$, and the estimation errors, $\mathcal C_{\theta}$ and $\mathcal C_{\eta}$, are small. The estimation errors $\mathcal{C}_\theta$ and $\mathcal{C}_\eta$ vanish as the data sizes $N_1$ and $N_2$ increase, which justifies our data-generation procedure. On the other hand, the approximation errors can be made small by using sufficiently rich network architectures $\mathcal{H}_{\theta}$ and $\mathcal{H}_{\eta}$ (e.g., deep and wide) to be able to represent the true maps $\mathcal{T}$ and $\mathcal{T}^*$, by leveraging the universal approximation property of neural networks. 
It is important to note the theoretical trade-off here: while restricting the hypothesis class to bounded-norm weights guarantees that the loss bounds $M_\eta$ and $M_\theta$ in Lemmas~\ref{lem:R-rec-bound} and \ref{lem:R-pinn-bound} remain finite, restricting the weights too severely can limit the expressiveness of the network, thereby increasing the irreducible approximation error terms $\hat{R}_{S_1}$ and $\hat{R}_{S_2}$. Therefore, a sufficiently rich architecture must be used so that the parameter-bounded hypothesis class still contains accurate approximations.

The amplification factor $2\ell_\eta^2$ in \eqref{eq:inv-error-RT*-bound-2} reflects a key geometric ill-conditioning in KKL observers. 
When the eigenvalues of $A$ have large negative real parts, the PDE \eqref{eq:pde} makes the forward map $\mathcal{T}$ highly contractive, mapping the compact set $\mathcal{X}$ into a very small image manifold $\mathcal{T}(\mathcal{X}) \subset \mathcal{Z}$. 
As a result, the true inverse $\mathcal{T}^*$ becomes extremely sensitive and needs a large Lipschitz constant to recover the original states from this contracted space. 
Since the neural network $\hat{\mathcal{T}}_\eta^*$ must approximate this same inverse, its Lipschitz constant $\ell_\eta$ must grow accordingly, which enlarges the error bound. 
This reveals a fundamental design trade-off: the spectrum of $A$ must balance fast exponential convergence against the conditioning of $\mathcal{T}$, and explicit Lipschitz regularization is required during training to control $\ell_\eta$.

The role of our physics-informed learning method is as follows. 
According to \eqref{eq:inv-error-RT*-bound-2}, minimizing the inverse map error $R_{\mathcal{T}^*}$ requires minimizing the forward map error $R_\mathcal{T}$ (Lemma~\ref{lem:R-eta-bound}). A standard supervised method with $\nu=0$ over an overparameterized hypothesis class $\mathcal{H}_\theta$ only enforces pointwise interpolation on the training set. This often yields local minima that fit the training points but may violate the system's continuous dynamics between them. Incorporating the PDE penalty as in \eqref{eq:emp-risk-theta} mitigates this issue. While this penalty does not formally reduce the pseudo-dimension $d_\theta$ of $\mathcal{H}_\theta$, it regularizes the empirical risk by penalizing the network's spatial derivatives to satisfy \eqref{eq:pde_residual} over a dense set of collocation points $S_{\text{pde}}$. This restricts the optimization process to parameter configurations that satisfy the PDE~\eqref{eq:pde} across the state space.

\section{Robustness of Learned KKL Observer}
\label{sec_robustness}

The neural networks $\hat{\mc T}_\theta$ and $\hat{\mc T}_\eta^*$ aproximate $\mc T$ and $\mc T^*$, respectively. 
Approximation errors affect the observer's estimation performance when we use neural networks instead of the original transformation maps. 
For any $z\in\mc Z$, the true inverse $\mc T^*(z)$ can be written as
\be \label{eq:inverse_T}
\mc T^*(z) = \hat{\mc T}_\eta^*(z) + \mc{E}_\eta^*(z)
\ee 
where $\mc{E}_\eta^*$ denotes the approximation error of the learned inverse map $\hat{\mc T}_\eta^*$. 
The learned KKL observer is given by
\be \label{eq:kkl_observer2}
    \colsep=2pt\ba{ccl}
        \dot{\hat z}(t) &=& A\hat z(t) + By(t), \quad \hat z(0) = \hat z_0 \\
        \hat x(t) &=& \hat{\mc T}_\eta^*(\hat z(t)).
    \ea
\ee
The estimate obtained by this observer is affected by the approximation error $\mc{E}_\eta^*(\hat z(t))$, which acts as an unknown noise
Additionally, the system model \eqref{eq:sys} is rarely perfect in practice, and underlying uncertainties can affect the estimation performance.
In this section, we provide robustness guarantees for the estimation error under both approximation error and system uncertainties. 

\subsection{Problem Setup and Assumptions}
\label{subsec:7.1}

Consider an uncertain nonlinear system
\begin{subequations}
    \label{eq:sys-noisy}
    \begin{align}
        \dot{x}(t) &= f(x(t)) + w(t), \quad x(0)=x_0 
        \label{eq:sys-state-noisy} \\
        y(t) &= h(x(t)) + v(t)
        \label{eq:sys-output-noisy}
    \end{align}
\end{subequations} 
where $w(t)\in\bb R^{n_x}$ and $v(t)\in\bb R^{n_y}$ denote the model uncertainty and the measurement noise, respectively.
For uncertain systems, the objective of the observer changes from steering the estimation error asymptotically to zero to achieving a bounded time-average steady-state error. In immersion-based designs like the KKL observer, the observer state $\hat{z}(t)$ evolves in a higher-dimensional ambient space. During the initial transient phase, $\hat{z}(t)$ generally lies outside the image manifold $\mathcal{T}(\mathcal{X})$. Because the left-inverse $\mathcal{T}^*$ is valid only for points within $\mathcal{T}(\mathcal{X})$, applying the learned inverse map $\hat{\mathcal{T}}_\eta^*$ to an arbitrary $\hat{z}(t) \notin \mathcal{T}(\mathcal{X})$ does not guarantee the preservation of the contraction bounds required for input-to-state stability of the estimation error. Specifically, the estimation error $\|\xi(t)\|$ may experience transient excursions that are independent of the initial error $\|\xi(0)\|$ until $\hat{z}(t)$ converges to a neighborhood of $\mathcal{T}(\mathcal{X})$. Therefore, rather than enforcing a strict envelope on the transient behavior, we aim to guarantee that the time-average steady-state estimation error satisfies
\begin{equation}
    \label{eq:pAG-requirement}
    \limsup_{T \to \infty} \frac{1}{T} \! \int_0^T \!\! \|\xi(t)\|^2 \mathrm{d}t \le \gamma(\max(\|w\|_{L_\infty},\|v\|_{L_\infty})) + c
\end{equation}
where $\gamma$ is a class $\mathcal{K}_\infty$ function and $c \ge 0$ captures the approximation error introduced by the neural network $\hat{\mathcal{T}}_\eta^*$. From \eqref{eq:pAG-requirement}, it is evident that in the absence of uncertainty $w(t)$ and noise $v(t)$, the ultimate time-average error is bounded by the learning approximation limit $c$. Furthermore, \eqref{eq:pAG-requirement} imposes a robustness criterion on the observer's long-term performance, ensuring a graceful degradation of the average state estimate as the magnitudes of the disturbances increase.

Let $x(t;x_0,w)$ denote the state of \eqref{eq:sys-state-noisy} at time $t\in\bb R_{\geq 0}$ initialized at $x(0)=x_0$ and driven by uncertainty $w_{[0,t]}$.
When $w\equiv 0$, the state trajectory is denoted by $x(t;x_0)$.

\begin{assume} \label{assume-bounded-uncertainty}
The model uncertainty $w(t)$ and the measurement noise $v(t)$ satisfy the following:
\begin{enumerate}[(i)]
    \item Boundedness: {It holds that $\|w\|_{L_\infty} < \ol w$ and $\|v\|_{L_\infty} < \ol v$, for some known $\ol w,\ol v>0$.}
    \label{assume-bounded-uncertainty-i}
    \item \label{assume-bounded-uncertainty-ii}
    Bounded effect of $w$: There exists a class $\mc K_\infty$ function $\psi$ such that, for every $t\in\bb{R}_{\geq 0}$,
    \[
    \|x(t;x_0,w)-x(t;x_0)\| \leq \psi(\|w_{[0,t]}\|_{L_\infty}).
    \]
    \item {Robust forward invariance: For any $x_0 \in \mathcal{X}$ and any $w(t)$ satisfying $\|w\|_{L_\infty} < \overline{w}$, the true state $x(t; x_0, w)$ remains in $\mathcal{X}$.}
    \label{assume-bounded-uncertainty-iii}
\end{enumerate}
\end{assume}

Assumption~\ref{assume-bounded-uncertainty}(\ref{assume-bounded-uncertainty-i}) is quite standard in robust state estimation \cite{chen2018}. Assumption~\ref{assume-bounded-uncertainty}(\ref{assume-bounded-uncertainty-ii}) requires that the deterministic system~\eqref{eq:sys} does a good job in describing the uncertain system~\eqref{eq:sys-noisy}. Providing such a guarantee is widely studied in system identification, and the reader is referred to \cite{milanese1996, mania2022, ljung1999, abudia2022} for more details. 
{Finally, Assumption~\ref{assume-bounded-uncertainty}(\ref{assume-bounded-uncertainty-iii}) ensures the uniform injectivity of $\mathcal{T}$ is maintained along the state trajectory $x(t;x_0,w)$.}

Define the error in the $z$-coordinate as $\tilde z(t)\eqdef z(t)-\hat z(t)$. Then, from \eqref{eq:sys_z} and \eqref{eq:kkl_observer2}, we have
\be \label{eq:error_zeta}
    \dot{\tilde z}(t) = A\tilde z(t) + B \sigma(t)
\ee
where 
$
    \sigma(t)=h(\bar x(t))-h(x(t))-v(t)
$
with $x(t)\eqdef x(t;x_0,w)$ and $\bar{x}(t)\eqdef x(t;x_0)$.
Note that $\sigma(t)\in\bb{R}^{n_y}$ remains bounded due to Assumption~\ref{assume-bounded-uncertainty}. That is, for every $x(t),\bar{x}(t)\in\mc X$ and $t\in\bb{R}_{\geq 0}$,
\begin{align} \label{eq:sigma_ineq}
\|\sigma(t)\| &\leq  \|h(\bar x(t))-h(x(t))\| + \|v(t)\| \notag \\
&\leq \ell_h \|\bar x(t)-x(t)\| + \|v(t)\| \notag \\
&\leq \ell_h \psi(\|w_{[0,t]}\|_{L_\infty}) + \|v(t)\| \notag \\
&\leq \ell_h \psi(\ol w) + \sqrt{n_y} \ol v
\end{align}
where $\ol w,\ol v$ are from Assumption~\ref{assume-bounded-uncertainty}(i) and $\ell_h<\infty$ with
\be \label{eq:Lip-h}
\ell_h = \sup_{x\in\mc X} \sigma_{\max}\left(\frac{\partial h}{\partial x}(x)\right)
\ee
because $h(\cdot)$ is smooth and $\mc X$ is a compact set.
In the last step of \eqref{eq:sigma_ineq}, we used
\[
\|v(t)\|\leq \sqrt{n_y}\|v(t)\|_\infty\leq \sqrt{n_y} \|v_{[0,t]}\|_{L_\infty} \leq \sqrt{n_y} \ol v.
\]

\subsection{Technical Lemmas}

Recall $z=\mc T(x)$, where
$
\dot{z}(t) = Az(t) + Bh(\Bar{x}(t))
$
with $\Bar{x}(t)\eqdef x(t;x_0)$ the noise-free state trajectory. Also, recall the dynamics of $\hat z(t)=\hat{\mc T}_\theta(\hat x(t))$ given in \eqref{eq:kkl_observer2}.

\begin{lem} \label{lemma:exp_ineq}
    Suppose $A$ is Hurwitz and diagonalizable with eigenvalue decomposition $A=V\Lambda V^{-1}$. Then,
    \[
        \|\exp(At)\| \leq \cond(V) e^{\lambda_{\min}(A)t}
    \]
    and
    \[
        \int_0^t \|\exp(A\tau) B \| \mathrm{d}\tau \leq \frac{\cond(V)}{|\lambda_{\min}(A)|} \|B\| (1-e^{\lambda_{\min}(A)t}).
    \]
\end{lem}

A similar result can be obtained when $A$ is not diagonalizable by using the Jordan form of $A$; see \cite[Appendix C.5]{sontag2013}. However, since diagonalizable matrices are dense in the space of square matrices and also since $A$ is a design matrix, the diagonalizability assumption is mild.

\begin{lem} \label{lemma:zeta_bound}
Let Assumption~\ref{assume-bounded-uncertainty} hold. Then, the error $\tilde z(t)\eqdef z(t)-\hat z(t)$ in the $z$-coordinate satisfies
\begin{equation}
    \label{eq:zeta_bound}
    \|\tilde z(t)\| \leq \|\tilde z_0\| \cond(V) e^{\lambda_{\min}(A)t}
    + \frac{\cond(V)}{|\lambda_{\min}(A)|} \|B\| (1-e^{\lambda_{\min}(A)t}) \big[\ell_h \psi(\ol w) + \sqrt{n_y} \ol v \big]
\end{equation}
where $\ol w, \ol v, \psi$ are given in Assumption~\ref{assume-bounded-uncertainty}, and $\ell_h$ is the Lipschitz constant of $h(x)$ for $x\in\mc X$ given in \eqref{eq:Lip-h}.
\end{lem}
\begin{proof}
    Recall $\sigma(t)= h(\bar{x}(t)) - h(x(t))-v(t)$ from \eqref{eq:error_zeta}. Then, the solution $\tilde z(t;\tilde z_0,\sigma)$ of \eqref{eq:error_zeta} satisfies
    \be
    \label{eq:ineq_ztilde}
    \|\tilde z(t)\| \!\leq\! \|\exp(At)\tilde z_0\| +\! \int_0^t \! \|\exp(A\tau)B\sigma(t-\tau)\| \mathrm{d}\tau.
    \ee
    The second term on the right-hand side of \eqref{eq:ineq_ztilde} satisfies
    \begin{align}
    \int_0^t \|\exp(A\tau)B\sigma(t-\tau)\| \mathrm{d}\tau 
    & \leq \int_0^t \|\exp(A\tau)B\| \|\sigma(t-\tau)\| \mathrm{d}\tau \nonumber \\
    & \leq \int_0^t \|\exp(A\tau)B\| \mathrm{d}\tau \, \max_{\tau\geq 0} \|\sigma(\tau)\|.
    \label{eq:ineq_2}
    \end{align}
    From \eqref{eq:sigma_ineq}, it follows that 
    $
    \max_{\tau\geq 0}\|\sigma(\tau)\| \leq \ell_h\psi(\ol w) + \sqrt{n_y}\ol v.
    $
    Substituting this in \eqref{eq:ineq_2} and in \eqref{eq:ineq_ztilde}, and using Lemma~\ref{lemma:exp_ineq}, we obtain \eqref{eq:zeta_bound}. 
\end{proof}

\subsection{Robustness to the Approximation Error} \label{subsec_robust_approx_error}

To analyze the observer's performance, we employ some measure-theoretic concepts. Recall the general probability measure $\mc D_{\mc X}$ over the compact state space $\mathcal{X}$. To describe the asymptotic, long-term statistical distribution of the system's trajectories, we consider physical invariant measures \cite{eckmann1985, young2002}. Given a system \eqref{eq:sys}, $\mu$ is a physical invariant measure if the probability of the state $x$ being in any set $X \subseteq \mathcal{X}$ remains constant under the system's flow \cite{ashwin2021}. Note that in deterministic dynamical systems, the state evolution itself is not random. 
However, the randomness originates from the system initialization. 
Therefore, any probabilistic statement refers to an ensemble of trajectories generated by a distribution of initial conditions.

The primary objective of the observer is to minimize estimation error under the system's long-term, steady-state operating conditions. For many nonlinear systems, especially the ones we consider in Section~\ref{sec_simulations}, the asymptotic behavior is confined to the attractor, whose statistics are governed by the invariant measure $\mu$. 
Consequently, we evaluate the observer's asymptotic performance using the time-average steady-state estimation error.

Given the invariant measure $\mu$, the injective KKL transformation map $\mathcal{T}: \mathcal{X} \to \mathcal{Z}$ induces a pushforward measure $\mu_{\mathcal{T}}$ on the observer space $\mathcal{Z}$, defined by $\mu_{\mathcal{T}}(Z) = \mu(\mathcal{T}^{-1}(Z))$, $\forall Z \subseteq \mathcal{Z}$, where $\mathcal{T}^{-1}(Z)$ is the preimage. The induced measure $\mu_{\mathcal{T}}$ represents the steady-state distribution of the transformed states $z$. 
Therefore, the expectation $\mathbb{E}_{z \sim \mu_{\mathcal{T}}}[\cdot]$ denotes the integral over the observer state space $\mathcal{Z}$ with respect to $\mu_{\mathcal{T}}$. To guarantee that the infinite-horizon time average of a single trajectory converges to this measure-theoretic expectation, we require the following standard assumption.

\begin{assume}
    \label{assume:ergodic}
    The physical invariant measure $\mu$ supported on the compact set $\mathcal{X}$ is an ergodic measure with respect to the flow generated by $f(x)$ in \eqref{eq:sys-state}.
\end{assume}

Under Assumption~\ref{assume:ergodic}, the pushforward measure $\mu_{\mathcal{T}}$ is ergodic for the steady-state trajectories of $z(t)$ on the manifold $\mathcal{T}(\mathcal{X})$. 
For systems exhibiting stable limit cycles (e.g., the Van der Pol oscillator), the invariant measure is strictly and uniquely ergodic. For systems exhibiting deterministic chaos (e.g., the Lorenz and Rössler attractors), it is a well-established result that they admit Sinai-Ruelle-Bowen (SRB) measures \cite{young2002}, which are ergodic physical invariant measures that characterize the time-average behavior of trajectories originating from a basin of attraction with positive Lebesgue measure.

Since the estimate $\hat{x}(t)$ is generated by the learned observer \eqref{eq:kkl_observer2}, the estimation error is given by $\xi(t) = x(t)-\hat{x}(t) = \mathcal{T}^*(z(t)) - \hat{\mathcal{T}}_\eta^*(\hat{z}(t))$.
Theorem~\ref{thm:R-eta-bound} provides a high-probability bound on the expected risk $R_{\mathcal{T}^*}$, ensuring that the neural network $\hat{\mathcal{T}}_\eta^*$ accurately approximates the true inverse $\mathcal{T}^*$ with respect to the measure $\mu_{\mathcal{T}}$. However, as established in Section~\ref{subsec:7.1}, this learning guarantee is only supported on the image manifold $\mathcal{T}(\mathcal{X})$. While the global Lipschitz continuity of the neural network (Assumption~\ref{assume-Lipschitz}) ensures that the estimate $\hat{x}(t)$ remains bounded for any arbitrary transient state $\hat{z}(t) \in \mathbb{R}^{n_z}$, the network lacks approximation guarantees when evaluating points outside $\mathcal{T}(\mathcal{X})$. Consequently, deriving tight, point-wise transient error bounds in the original coordinates is mathematically unfeasible. Therefore, we evaluate the observer performance by showing how $R_{\mathcal{T}^*}$ directly bounds the infinite-horizon time-average steady-state estimation error. We first analyze the noise-free case and then extend the result to uncertain systems.

\begin{thm}
\label{thm:robust-approx}
Let Assumptions~\ref{assumption_1}, \ref{assumption_2}, \ref{assume-Lipschitz}, and \ref{assume:ergodic} hold. Let $\hat x(t)$ be the state estimate from the learned KKL observer \eqref{eq:kkl_observer2} for the noise-free system \eqref{eq:sys}. Then, the time-average steady-state estimation error is bounded as
\begin{equation*}
    {\limsup_{T \to \infty} \frac{1}{T} \int_0^T \|\xi(t)\|^2 \mathrm{d}t \le 2R_{\mathcal{T}^*}}
\end{equation*}
where $R_{\mc T^*}$ is the inverse map error \eqref{eq:inv-error-RT*} bounded by \eqref{eq:inv-error-RT*-bound-2}.
\end{thm}
\begin{proof}
    From \eqref{eq:inverse_T} and \eqref{eq:error_zeta}, the estimation error $\xi(t) = x(t) - \hat{x}(t)$ satisfies
    \begin{align*}
    \|\xi(t)\| &= \|\mathcal{T}^*(z(t)) - \hat{\mathcal{T}}_{\eta}^*(\hat{z}(t))\| \\
    &\le \| \hat{\mc T}_\eta^*(z(t)) - \hat{\mc T}_\eta^*(\hat z(t)) \| + \| \mc T^*(z(t)) - \hat{\mc T}_\eta^*(z(t)) \| \\
    &\le \ell_{\eta}\|\tilde{z}(t)\| + \|\mc E_\eta^*(z(t))\|
    \end{align*}
    where $\tilde{z}(t) = \hat{z}(t) - z(t)$ and $\mc E_\eta^*(z(t))$ given by \eqref{eq:inverse_T}. Using the inequality $(a+b)^2 \le 2a^2 + 2b^2$, we have
    $$
    \|\xi(t)\|^2 \le 2\ell_{\eta}^2 \|\tilde{z}(t)\|^2 + 2 \|\mc E_\eta^*(z(t))\|^2.
    $$
    Integrating over time and taking the superior limit yields
    \begin{equation}
        \label{eq:ineq-thm18-pf}
        \limsup_{T \to \infty} \frac{1}{T} \int_0^T \|\xi(t)\|^2 \mathrm{d}t \le \limsup_{T \to \infty} \frac{2\ell_\eta^2}{T} \int_0^T \|\tilde{z}(t)\|^2 \mathrm{d}t
        + \limsup_{T \to \infty} \frac{2}{T} \int_0^T \|\mc E_\eta^*(z(t))\|^2 \mathrm{d}t.
    \end{equation}
    Since $A$ is Hurwitz, $\tilde{z}(t)$ converges to zero exponentially. Thus, the integral in the first term is bounded, and the limit equals zero.
    For the second term, by Birkhoff's Ergodic Theorem \cite[Th. 4.2.4.]{lasota2013}, the time-average of the error along the trajectory converges to its spatial expectation over $\mu_{\mathcal{T}}$, i.e., 
    \begin{equation}
    \label{eq:ineq-thm18-pf2}
        \lim_{T \to \infty} \frac{1}{T} \int_0^T \|\mc E_\eta^*(z(t))\|^2 \mathrm{d}t = \mathbb{E}_{z \sim \mu_{\mathcal{T}}} [\|\mc E_\eta^*(z)\|^2] = R_{\mc T^*}.
    \end{equation}
    This concludes the proof.
\end{proof}

This theorem provides a link between the learning guarantee in Theorem~\ref{thm:R-eta-bound} and the observer performance. 
It shows that while the observer may exhibit persistent small-signal fluctuations around the true state due to neural network approximation error, these deviations remain uniformly bounded in magnitude. 
As time passes, the observer converges to, and permanently remains within, a tight neighborhood of the true state trajectory, with the thickness of that neighborhood governed directly by the inverse map error $R_{\mc T^*}$.
This result is highly practical as Theorem~\ref{thm:R-eta-bound} gives a recipe for reducing $R_{\mc T^*}$ (i.e., more data, appropriate network complexity, $\ell_\eta$ regularization), whereas Theorem~\ref{thm:robust-approx} proves that doing so provably improves the average-case performance of the KKL observer's state estimate.

\subsection{Robustness to Uncertainty and Noise}

We analyze the robustness of the learned observer~\eqref{eq:kkl_observer2} to estimate the state of an uncertain system~\eqref{eq:sys-noisy}. Note that the design method of KKL observers as presented in Sections~\ref{sec_prelim} and \ref{sec_learning} remains the same for \eqref{eq:sys-noisy}.

\begin{thm} \label{thm:robust_noise}
    Let Assumptions~\ref{assumption_1}, \ref{assumption_2}, \ref{assume-Lipschitz}, \ref{assume-bounded-uncertainty}, and \ref{assume:ergodic} hold. 
    Let $\hat{x}(t)=\hat{\mc T}_\eta^*(\hat z(t))$ be the estimate from \eqref{eq:kkl_observer2} for the uncertain system \eqref{eq:sys-noisy}. Suppose $A$ is diagonalizable with eigendecomposition $A=V\Lambda V^{-1}$. Then, the time-average steady-state estimation error satisfies
    \begin{equation} 
    \label{eq:error_bound_2}
        \limsup_{T \to \infty} \frac{1}{T} \int_0^T \|\xi(t)\|^2 \mathrm{d}t \le 2R_{\mathcal{T}^*} \\
        + 2\ell_\eta^2 \left( \frac{\cond(V)}{|\lambda_{\min}(A)|} \|B\| \right)^2 \left( \ell_h \psi(\overline{w}) + \sqrt{n_y}\overline{v} \right)^2
    \end{equation}
    where $R_{\mc T^*}$ is bounded by \eqref{eq:inv-error-RT*-bound-2}, $\ol w,\ol v,\psi$ are given in Assumption~\ref{assume-bounded-uncertainty}, $\ell_h$ is given in \eqref{eq:Lip-h}, $\ell_\eta$ is given in \eqref{eq:inverse_T_lipschitz}, and $n_y$ is the dimension of the output vector $y(t)$.
\end{thm}
\begin{proof}
    We begin from \eqref{eq:ineq-thm18-pf} in the proof of Theorem~\ref{thm:robust-approx}, where the second term on the right-hand side is bounded by \eqref{eq:ineq-thm18-pf2}. From Lemma~\ref{lemma:zeta_bound}, we have
    $$
    \limsup_{t \to \infty} \|\tilde{z}(t)\| \le \frac{\cond(V)}{|\lambda_{\min}(A)|} \|B\| \left( \ell_h \psi(\overline{w}) + \sqrt{n_y}\overline{v} \right) =: \tilde{\mathcal B}.
    $$
    By the definition of the limit superior, for any arbitrarily small $\epsilon > 0$, there exists a finite time $t_0 > 0$ such that for all $t \ge t_0$, $\|\tilde{z}(t)\|^2 \le (\tilde{\mathcal B} + \epsilon)^2$. Splitting the integral (first term on the right-hand side of \eqref{eq:ineq-thm18-pf}), we obtain
    \begin{align*}
    \frac{1}{T} \int_0^T \|\tilde{z}(t)\|^2 \mathrm{d}t &= \frac{1}{T} \int_0^{t_0} \|\tilde{z}(t)\|^2 \mathrm{d}t + \frac{1}{T} \int_{t_0}^T \|\tilde{z}(t)\|^2 \mathrm{d}t \\
    & \le \frac{1}{T} \int_0^{t_0} \|\tilde{z}(t)\|^2 \mathrm{d}t + \frac{T - t_0}{T} (\tilde{\mathcal B} + \epsilon)^2.
    \end{align*}
    As $T \to \infty$, the first term vanishes because the integral is bounded, and $\frac{T-t_0}{T} \to 1$. Since the above inequality holds for any $\epsilon > 0$, we take the limit as $\epsilon \to 0$ to obtain
    $$
    \limsup_{T \to \infty} \frac{1}{T} \int_0^T \|\tilde{z}(t)\|^2 \mathrm{d}t \le \tilde{\mathcal B}^2.
    $$
    Substituting this result and \eqref{eq:ineq-thm18-pf2} back into the time-average inequality \eqref{eq:ineq-thm18-pf} yields the final bound \eqref{eq:error_bound_2}. This concludes the proof.
\end{proof}

This theorem shows that the time-average steady-state error is additively composed of two terms: a learning-dependent term and a design-dependent term. The learning-dependent term $R_{\mc T^*}$ is the approximation error for the learned inverse map $\hat{\mc T}_\eta^*$, which we can control via the learning process (Theorem~\ref{thm:R-eta-bound}). The design-dependent term depends on the noise bounds ($\overline{w}, \overline{v}$) and the observer's design parameters ($A, B, \ell_\eta$). This term can be minimized by choosing $A$ to be stable ($|\lambda_{\min}(A)|$ large) and well-conditioned ($\cond(V)$ small\footnote{For instance, when $A$ is a diagonal matrix, we have $V=I_{n_z}$ and $\cond(V)=1$.}), and by regularizing the learning of $\hat{\mathcal{T}}_{\eta}^*$ to keep $\ell_\eta$ small \cite{goodfellow2016}.

\section{Simulation Results}
\label{sec_simulations}

In this section, we demonstrate the effectiveness of our proposed methodology\footnote{Our code: \href{https://github.com/Mudhdhoo/PINN_KKL}{\url{github.com/Mudhdhoo/PINN_KKL}} and \href{https://github.com/mubniazi/spinn-KKL}{\url{github.com/mubniazi/spinn-KKL}}} for the learning-based design of KKL observers using the following examples:
\begin{itemize}
    \item \textit{Reverse Duffing oscillator}:
    \begin{equation}\label{eq:rev_duff}
        \dot{x}_1 = x_2^3, \quad \dot{x}_2 = -x_1, \quad 
        y = x_1.
    \end{equation}
    
    \item \textit{Van der Pol oscillator} with $\mu = 3$:
    \begin{equation}\label{eq:vdp}
        \dot{x}_1 = x_2, \quad \dot{x}_2 = \mu(1-x_1^2)x_2 - x_1, \quad
        y = x_1.
    \end{equation}
    
    \item \textit{R\"{o}ssler attractor} with $a\!=\!b\!=\!0.2$, and $c\!=\!5.7$:
    \begin{equation}\label{eq:ross}
        \dot{x}_1 = -x_2 - x_3, \quad \dot{x}_2 = x_1 + ax_2, \quad
        \dot{x}_3 = b + x_3(x_1 - c), \quad y = x_2.
    \end{equation}
    
    \item \textit{Lorenz attractor} with $p=28$, $q=10$, $r=8/3$:
    \begin{equation}\label{eq:lorenz}
        \dot{x}_1 = p(x_2 - x_1), \quad
        \dot{x}_2 = x_1(q - x_3) - x_2, \quad
        \dot{x}_3 = x_1 x_2 - r x_3, \quad
        y = x_2.
    \end{equation}
\end{itemize}

\begin{figure}[!ht]
    \centering
    \begin{subfigure}{0.7\textwidth}
        \includegraphics[width=\textwidth]{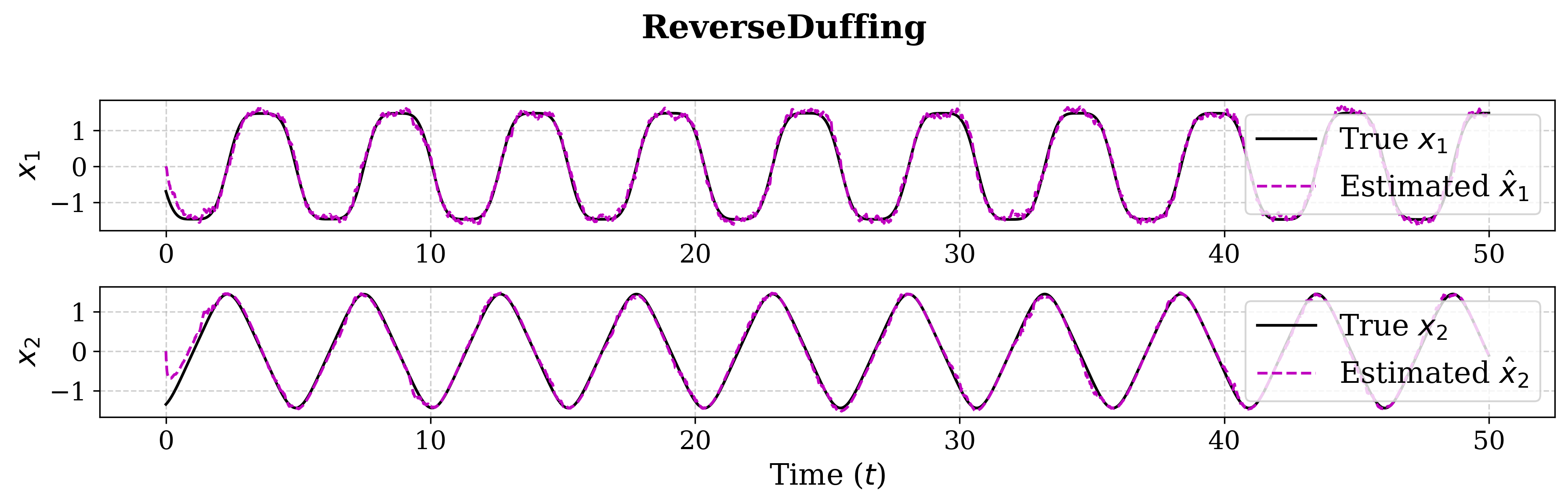}
    \end{subfigure}

    \begin{subfigure}{0.7\textwidth}
        \includegraphics[width=\textwidth]{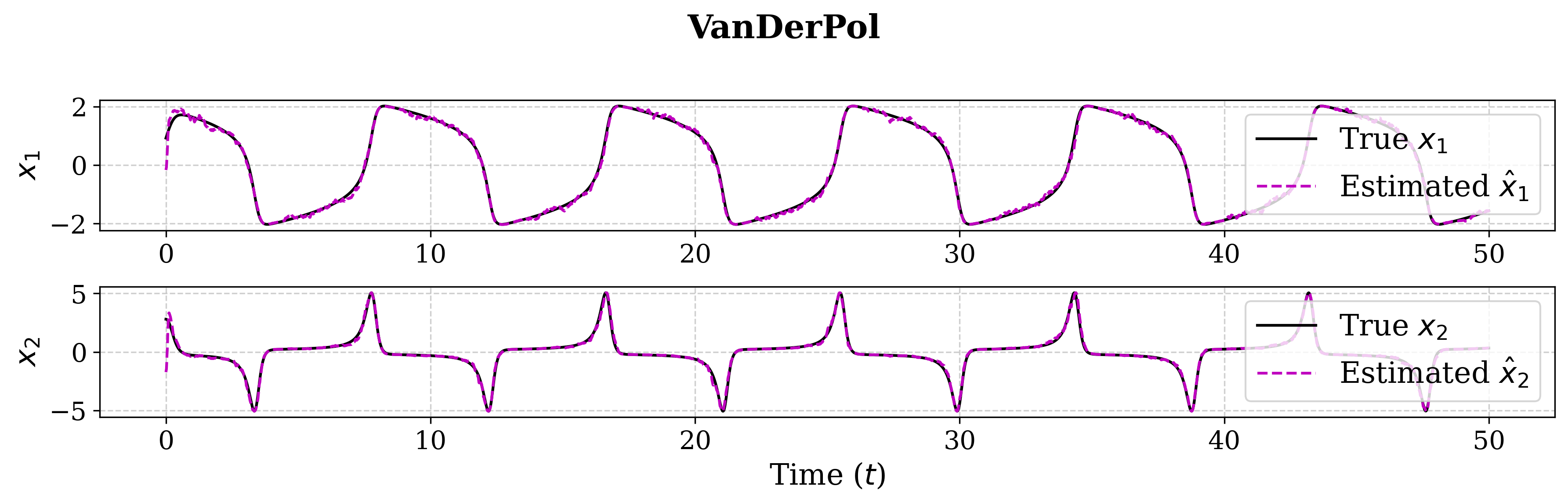}
    \end{subfigure}

    \begin{subfigure}{0.7\textwidth}
        \includegraphics[width=\textwidth]{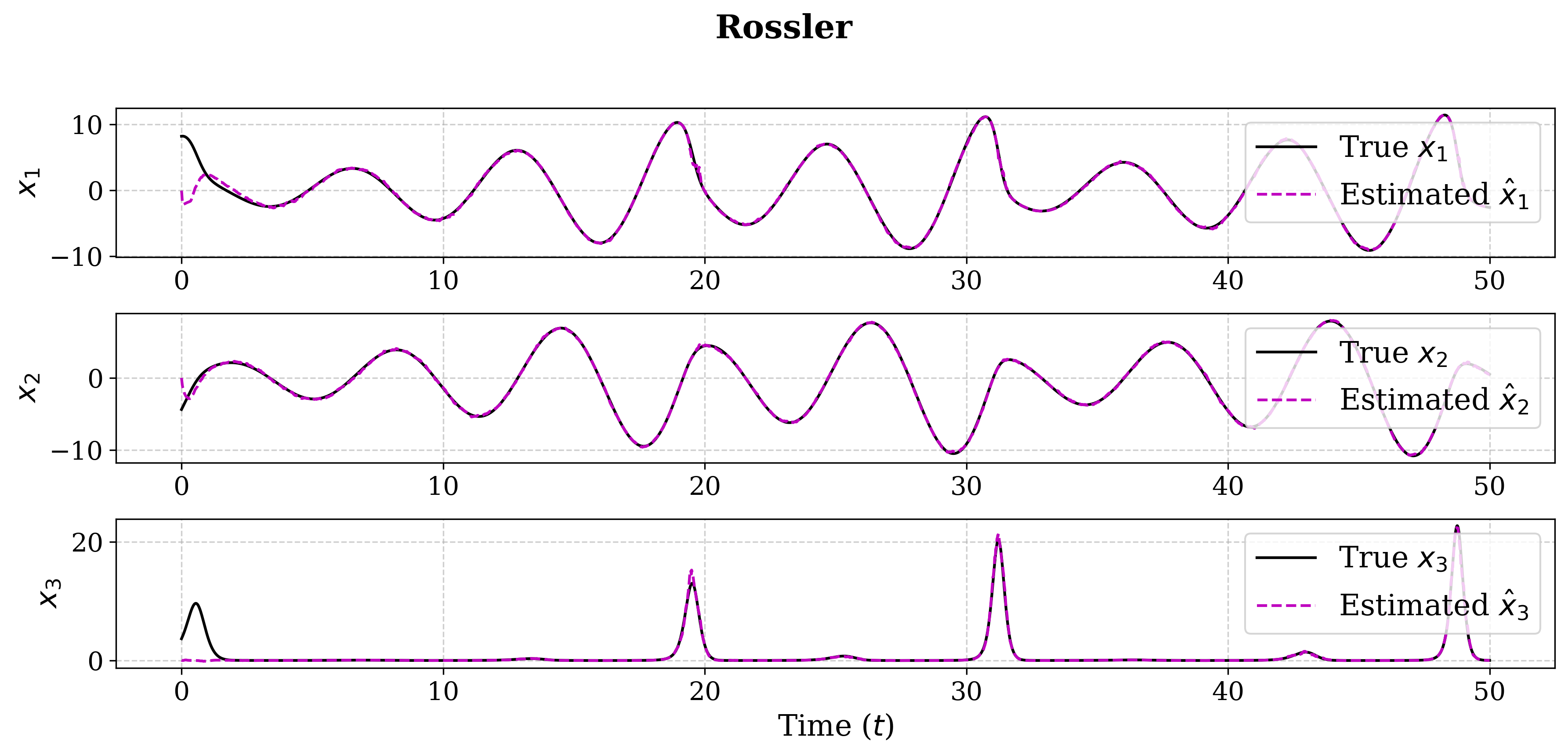}
    \end{subfigure}

    \begin{subfigure}{0.7\textwidth}
        \includegraphics[width=\textwidth]{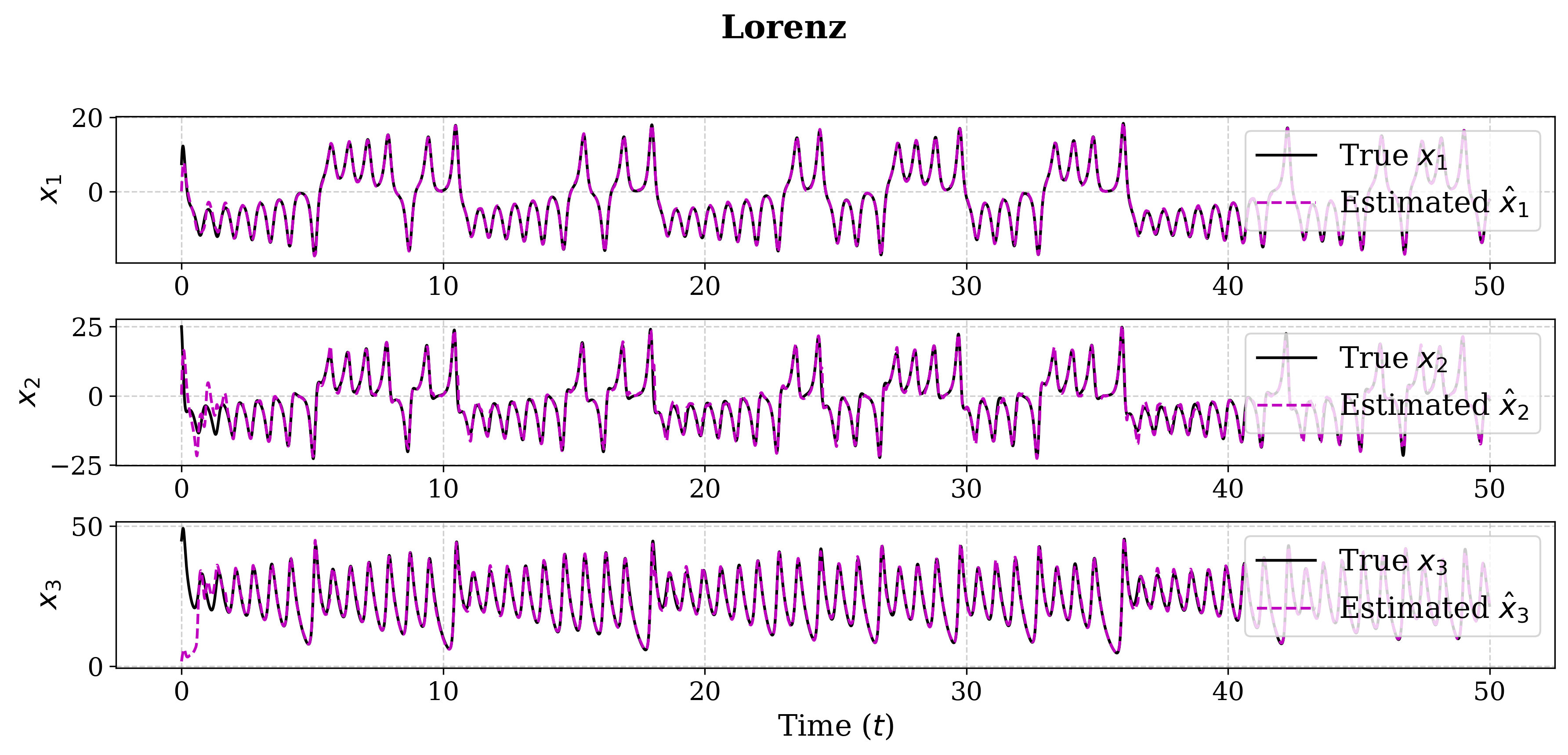}
    \end{subfigure}
    
    \caption{State estimation with learned KKL observer under noisy measurements.
    The solid black line shows the true state, while the dashed magenta line is the estimate obtained from the learned KKL observer.}
    \label{fig:sim_results}
\end{figure}

\subsection{Experimental Setup} 
\label{subsec_exp-setup}

We generate synthetic datasets $S_1=\{S_\text{data},S_\text{pde}\}$ and $S_2$ according to the procedure in Section~\ref{subsec_datagen}.
For all systems, we uniformly sample 500 initial conditions in $\mathcal{X}_\text{train}$, which equals 
\begin{align*}
    \mathcal{X}_\text{train} &= \{[-2,2],[-2,2]\} \\
    \mathcal{X}_\text{train} &= \{[-5,5],[-8,8]\} \\
    \mathcal{X}_\text{train} &= \{[-10,10],[-10,10],[0,20]\} \\
    \mathcal{X}_\text{train} &= \{[-20,20],[-30,30],[0,50]\}
\end{align*}
for \eqref{eq:rev_duff}--\eqref{eq:lorenz}, respectively.
From these initial conditions, we generate $x$ and $z$ trajectories for $S_\text{data}$ over the time interval $[0, 50]$ with a sampling time of $0.01$. The truncation sample $k_*$ is set to $0.2\tau$ (20\% of the $z$-trajectory length), which is sufficient to ensure that the effects of all initial conditions vanish by $\varepsilon = 10^{-6}$. This gives $|S_\text{data}|=2\times 10^6$ points.
We also sample $10^6$ points for $S_\text{pde}$ from $\mathcal{X}_\text{train}$.
Thus, $|S_1|=3\times 10^6$ points.
Notice that the data points $S_\text{data}$ and the collocation points $S_\text{pde}$ are sampled independently. Therefore, they are almost surely disjoint ($S_\text{data} \cap S_\text{pde} = \emptyset$ a.s.). We sample $5\times 10^6$ points from $\mathcal{X}_\text{train}$ for $S_2$. The observer matrices $A=\mathrm{diag}(\lambda_1,\dots,\lambda_{n_z})$, where $\lambda_i\in[-3,-1.5]$ are distinct, and $B=\mathbf{1}_{n_z}$. For reverse Duffing and Van der Pol, $n_z=5$, and for Rössler and Lorenz, $n_z=7$.

We model the forward map $\hat{\mathcal{T}}_\theta$ and the inverse map $\hat{\mathcal{T}}_\eta$ as multilayer {perceptrons}, trained on the synthetic datasets generated for \eqref{eq:rev_duff}--\eqref{eq:lorenz} according to the methodology in Section~\ref{sec_learning}. The hyperparameter specifications for each model are presented in Table~\ref{tab:hyperparams}.

\begin{table}[!hb]
    \centering
    \caption{Hyperparameters used for training. All values are used for both the forward map $\hat{\mc T}_\theta$ and the inverse map $\hat{\mc T}_\eta$.}
    \label{tab:hyperparams}
    \renewcommand{\arraystretch}{1.4} 
    \setlength{\tabcolsep}{3pt}
    \begin{tabular}{ P{0.15\linewidth} P{0.14\linewidth} P{0.1\linewidth} P{0.15\linewidth} P{0.15\linewidth} P{0.11\linewidth} } 
        \toprule
        \textbf{Activation function} & \textbf{{Hidden} Layers} & \textbf{Layer Size} & \textbf{Learning Rate} & {\textbf{Physics weight~$\nu$}} & \textbf{Number of Epochs} \\
        \midrule
        SiLU$(\cdot)$ & 4 & 512 & $1\times 10^{-3}$ & 0.1 & 200 \\ 
        \bottomrule
    \end{tabular}
\end{table}

\subsection{Estimation Results under Measurement Noise}

Simulation results for each system are illustrated in Fig.~\ref{fig:sim_results}. Each trajectory is generated from random initial condition, different from those in the dataset $S_1$ but sampled within the same region {$\mathcal{X}_\text{train}$}. Each component of the measurement noise is sampled from {$\mathcal{N}(0, 0.5^2)$} for \eqref{eq:rev_duff}--\eqref{eq:lorenz}. 
The simulations demonstrate our method's ability to estimate the state with remarkable accuracy, even in the presence of measurement noise.

\subsection{Comparison with the State-of-the-Art}

We compare the state estimation performance of our sequential learning PINN method (referred to as SPINN) with three established learning-based approaches:
\begin{itemize}
    \item The neural ODE method (NODE) proposed by \cite{miao2023}. This method simultaneously learns the system model and the KKL observer using a neural ODE.\footnote{Code: \href{https://github.com/KYMiao/L4DC_Neural_ODEs_Observer}{\url{github.com/KYMiao/L4DC_Neural_ODEs_Observer}}}
    
    \item The deep model-free switching method (DMF) proposed by \cite{peralez2024}. This method learns the KKL observer only from the data, without relying on the model.\footnote{Code: \href{https://github.com/jolindien-git/DeepKKL/tree/main/L4DC_2024}{\url{github.com/jolindien-git/DeepKKL/tree/main/L4DC_2024}}}
    
    \item The unsupervised autoencoder method (AE) by \cite{buisson2023}. This is a joint encoder-decoder trained in an unsupervised way, with tuning of the observer gain.\footnote{Code: \href{https://github.com/Centre-automatique-et-systemes/learn_observe_KKL}{\url{github.com/Centre-automatique-et-systemes/learn_observe_KKL}}.}
\end{itemize}

The benchmark systems are \eqref{eq:rev_duff}--\eqref{eq:lorenz}.
For all methods, we use the same hyperparameters as in Table~\ref{tab:hyperparams} to ensure a fair comparison and focus solely on the differences in the learning method.
Moreover, we enforce a data budget across all methods with $N_\text{baseline} = |S_1|+|S_2| = 8\times 10^6$ data points, which matches our setup in Section~\ref{subsec_exp-setup}. Thus, the training of NODE, DMF, and AE is exposed to the same number of data points drawn from $\mathcal{X}_\text{train}$. The observed improvements in out-of-domain generalization are therefore due solely to the physics-informed learning, not to higher sampling density in $\mathcal{X}_\text{train}$.

In this comparison, we evaluate the performance and generalization capabilities of the learned observers produced by SPINN (ours), NODE, DMF, and AE. 
All methods are trained on data generated in $\mathcal{X}_\text{train}$ defined in Section~\ref{subsec_exp-setup}. The estimation performance is evaluated by testing on 100 independent trials with initial states of systems \eqref{eq:rev_duff}--\eqref{eq:lorenz} sampled from the test domains: (i)~$\mathcal{X}_\text{test}=\mathcal{X}_\text{train}$ and (ii)~$\mathcal{X}_\text{test}=2\mathcal{X}_\text{train}\setminus \mathcal{X}_\text{train}$.
In both cases, the data is generated independently.
We quantify the estimation accuracy over the $N_\text{traj}=100$ independent test trajectories via the symmetric mean absolute percentage error (SMAPE) for each trajectory $j$:
\[
\text{SMAPE}_j = \frac{1}{N_\text{steps}} \! \sum_{k=1}^{N_\text{steps}} \! \frac{2\|x_j(t_k) - \hat{x}_j(t_k)\|}{\|x_j(t_k)\| + \|\hat{x}_j(t_k)\|} \times 100\%
\]
which measures the relative estimation error across different scales.
{Because chaotic nonlinear systems can produce heavy‑tailed error distributions, the arithmetic mean over $N_\text{traj}$ trajectories is highly sensitive to rare catastrophic outliers. Instead, we report the $\text{Median}(\{\text{SMAPE}_j\}_{j=1}^{N_\mathrm{traj}})$ to better capture typical performance, together with the interquartile range (IQR) to quantify the spread and reliability of the observer's predictions across all trials.}

\begin{figure}[!ht]
    \centering
    \begin{subfigure}{0.7\textwidth}
        \includegraphics[width=\linewidth]{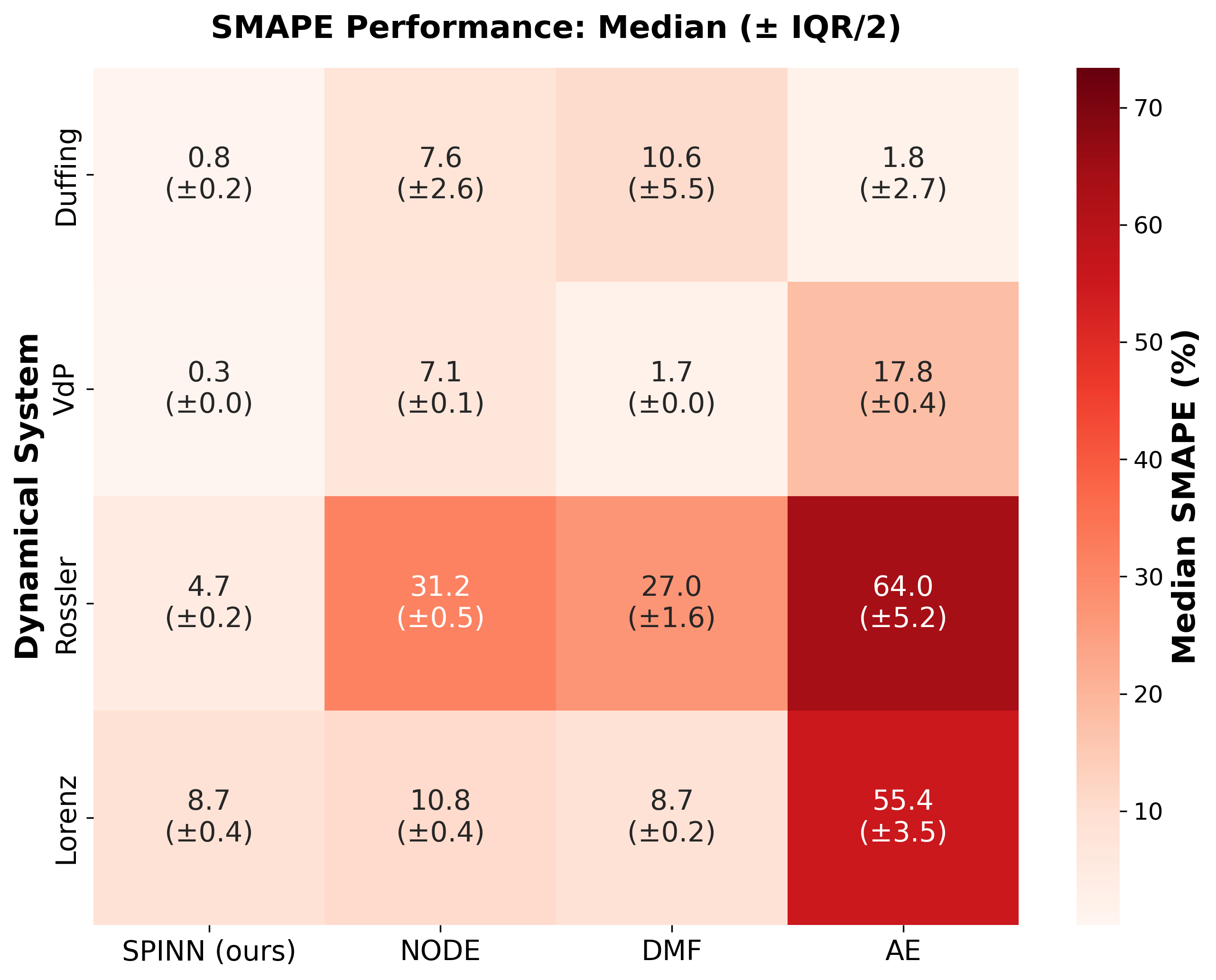}
        \caption{In-domain validation: $\mathcal{X}_\text{test}=\mathcal{X}_\text{train}$.}
    \end{subfigure}

    \vspace{6pt}
    \begin{subfigure}{0.7\textwidth}
        \includegraphics[width=\linewidth]{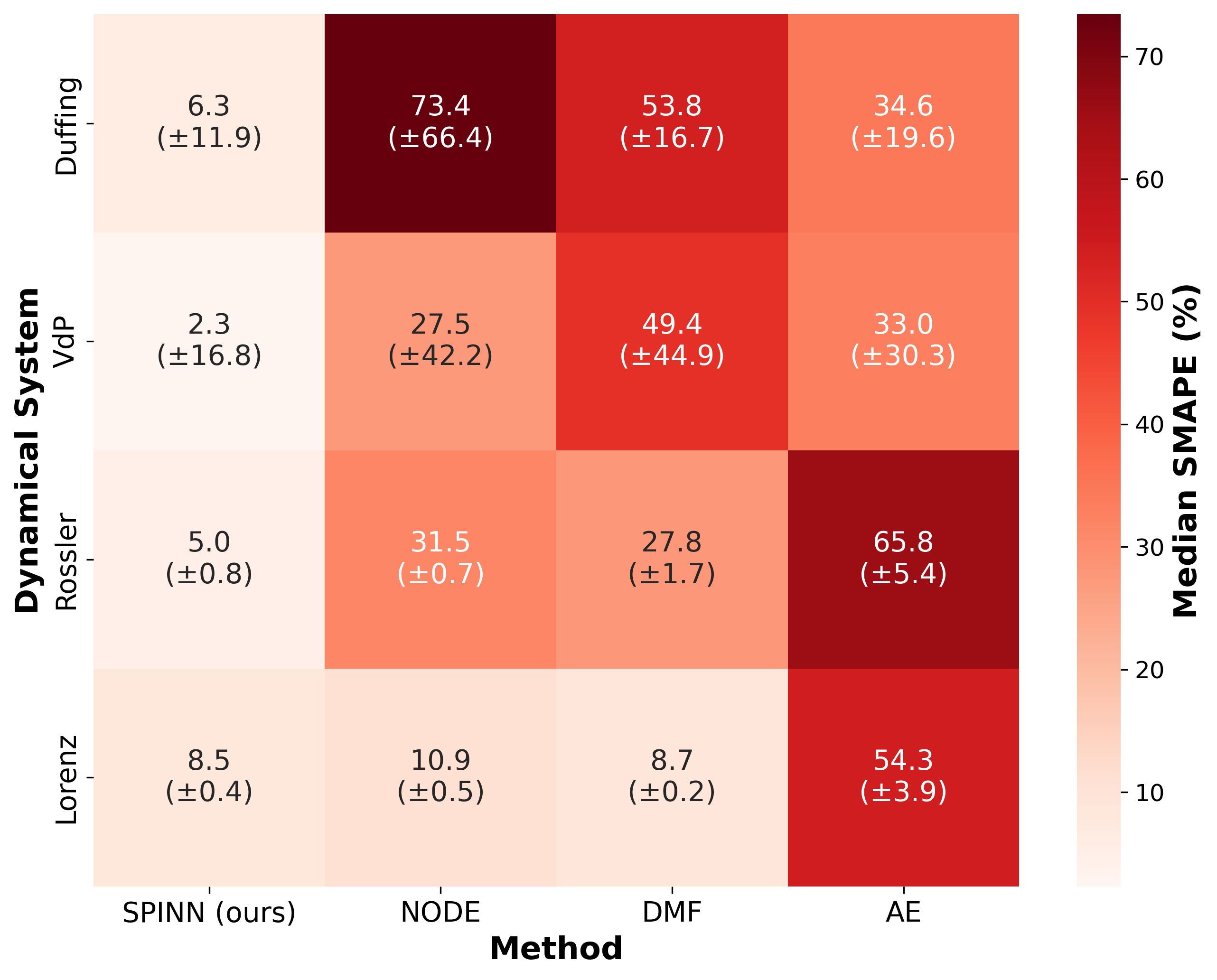}
        \caption{Out-of-domain validation: $\mathcal{X}_\text{test}=2\mathcal{X}_\text{train}\setminus \mathcal{X}_\text{train}$.}
    \end{subfigure}
    \caption{{Performance comparison of KKL observer synthesis methods across \eqref{eq:rev_duff}--\eqref{eq:lorenz}. Cell colors correspond to the median trajectory-wise SMAPE evaluated over 100 independent trials initialized in the test domain $\mathcal{X}_\text{test}$.}}
    \label{fig:kkl-comparison}
\end{figure}

Fig.~\ref{fig:kkl-comparison} shows that our method consistently achieves the lowest SMAPE across \eqref{eq:rev_duff}--\eqref{eq:lorenz}, indicating better performance and generalization in both in-domain and out-of-domain validation.
The AE method produced the largest errors on all benchmarks except reverse Duffing. NODE and DMF also degrade sharply in the out-of-domain validation for the reverse Duffing and Van der Pol oscillators, both of which have a single attracting limit cycle. 
Because these methods are purely data-driven and do not incorporate the governing differential equations, they lack the structure needed to infer how unseen states converge to the limit cycle when initialized outside the training distribution.

In contrast, SPINN avoids this failure mode by simply enforcing the PDE constraint for the KKL transformation directly. This physics constraint acts as an inductive bias that narrows the hypothesis space of the neural network. As a result, the learned mappings preserve key invariants, such as the Van der Pol limit cycle, and extrapolate reliably to larger, high-energy regions where purely data-driven methods break down.

Finally, we note that NODE and DMF generalize better on the Lorenz system because trajectory unrolling captures its dissipative dynamics. Temporal integration during training embeds the system’s inward contraction toward the attractor, helping the learned model recover off-manifold states. In contrast, AE fails under extrapolation because it learns only a static reconstruction manifold and does not model the surrounding vector field.

\section{Conclusion}
\label{sec_conclusion}

We proposed a learning-based method for designing KKL observers for autonomous nonlinear systems. We trained neural networks using synthetically generated data to approximate the transformation map and its inverse, which are required for synthesizing a KKL observer. Instead of simultaneously learning the transformation and its inverse, we proposed learning them sequentially to avoid conflicting gradients during training and improve accuracy. First, the transformation map is learned by a physics-informed neural network that incorporates the PDE associated with the KKL observer via automatic differentiation. {Then, by generating new data and using the learned transformation map to produce corresponding features, a second neural network learns the inverse map by minimizing the empirical reconstruction risk.}

We provided non-asymptotic generalization bounds for the learned KKL observer. We proved that minimizing the empirical risk {preserves the asymptotic robustness} of the state estimation error to approximation errors and system uncertainties.
The effectiveness of our approach was validated through comprehensive state estimation experiments across multiple benchmark systems. 
We demonstrated the generalization capability of the learned KKL observer by comparing it with the state-of-the-art methods. 
The better performance is because the PDE constraint serves as a physically consistent inductive bias that effectively reduces the hypothesis space for neural networks, enabling extrapolation to unseen data where purely data-driven methods fail.

Our future research will address some key open problems. The primary challenge lies in extending KKL observer learning to non-autonomous and controlled nonlinear systems, where the transformation maps and their inverses are time-varying. This temporal dependency results in time-dependent PDEs that probably require more complex neural network architectures. 
Moreover, combining PINNs for system identification with observer design would address the unknown system model scenario more robustly than current end-to-end learning methods,  which we used as baselines for comparison.
Finally, we will also explore active learning to identify where in the state space the PDE residual is high and sample adaptively from those regions, further reducing the data requirement.

\section*{Acknowledgments}
This work is supported by the European Union’s Horizon Research and Innovation Programme under Marie Skłodowska-Curie grant agreement No. 101062523. It also received funding from the Swedish Research Council's Distinguished Professor Grant, the Knut and Alice Wallenberg Foundation's Wallenberg Scholar Grant, and Wallenberg AI, Autonomous Systems and Software Program (WASP) funded by the Knut and Alice Wallenberg Foundation.

\bibliographystyle{alpha}
\bibliography{biblio}

\end{document}